\documentclass{aa}
\usepackage{astron}
\usepackage{graphics}
\usepackage{psfig}
\usepackage{times}
\def\etal{et al.}
\def\hii{H{\sc ii}}
\def\cbeta{$c_{\rm H\beta}$}

\def\ilam{$I_{\lambda}$}

\def\kms{km s$^{-1}$}
\def\kmsmpc{km s$^{-1}$ Mpc$^{-1}$}
\def\cmc{cm$^{-3}$}

\def\ergs{ergs s$^{-1}$}
\def\ergscm{ergs s$^{-1}$ cm$^{-2}$}

\def\msun{M$_{\odot}$}
\def\zsun{Z$_{\odot}$}
\def\halpha{\ifmmode {\rm H{\alpha}} \else $\rm H{\alpha}$\fi}
\def\hbeta{\ifmmode {\rm H{\beta}} \else $\rm H{\beta}$\fi}
\def\heia{He\,{\sc i} $\lambda$4471}
\def\heib{He\,{\sc i} $\lambda$4922}
\def\heic{He\,{\sc i} $\lambda$5876}
\def\heiia{He\,{\sc ii} $\lambda$4686}
\def\heiib{He\,{\sc ii} $\lambda$5412}
\def\oia{[O\,{\sc i}] $\lambda$6300}
\def\oib{[O\,{\sc i}] $\lambda$6364}
\def\oii{[O\,{\sc ii}] $\lambda$3727}
\def\oiii{[O\,{\sc iii}]}
\def\oiiia{[O\,{\sc iii}] $\lambda$4959}
\def\oiiib{[O\,{\sc iii}] $\lambda$5007}
\def\ov{O\,{\sc v} $\lambda$5590}
\def\ni{[N\,{\sc i}] $\lambda$5199}
\def\nii{[N\,{\sc ii}] $\lambda$5755}
\def\niii{N\,{\sc iii} $\lambda$4640}
\def\nv{N\,{\sc v} $\lambda$4604}
\def\Ciii{C\,{\sc iii} $\lambda$4650}
\def\ciii{C\,{\sc iii} $\lambda$5696}
\def\civ{C\,{\sc iv} $\lambda$5808}
\def\siii{[S\,{\sc iii}] $\lambda$6318}
\def\ariva{[Ar\,{\sc iv}] $\lambda$4711}
\def\arivb{[Ar\,{\sc iv}] $\lambda$4740}
\def\feiiia{[Fe\,{\sc iii}] $\lambda$4658}
\def\feiiib{[Fe\,{\sc iii}] $\lambda$5271}
\def\cliiia{[Cl\,{\sc iii}] $\lambda$5518}
\def\cliiib{[Cl\,{\sc iii}] $\lambda$5538}
\begin{document}

   \thesaurus{} 
   \title{Populations of WC and WN stars in Wolf-Rayet galaxies 
\thanks{Based on observations obtained at the European Southern Observatory (ESO), 
Chile}}

   \author{Daniel Schaerer \inst{1,2} \and
       	   Thierry Contini  \inst{3} 
	\thanks{{\it Present address: }European Southern Observatory, 
		Karl-Schwarzschild-Str. 2, D-85748 Garching bei M\"unchen, Germany}
	 \and
	   Daniel Kunth    \inst{4} 
	  }

   \offprints{D. Schaerer, schaerer@obs-mip.fr}


   \institute{
Observatoire Midi-Pyr\'en\'ees, Laboratoire d'Astrophysique, 14, Av. E. Belin, 
F-31400 Toulouse, France
\and
Space Telescope Science Institute, 3700 San Martin Drive, Baltimore, MD 21218, 
USA
\and
School of Physics \& Astronomy, Tel Aviv University, 69978 Tel Aviv, Israel
\and 
Institut d'Astrophysique de Paris, 98 bis Bd. Arago, F-75014 Paris, France
             }
   \date{Received 3 September 1998; accepted 25 Septembre 1998}

   \maketitle

   \begin{abstract} 
We report the detection of WC stars in five Wolf-Rayet (W-R) galaxies: 
He 2-10, NGC 3049, NGC 3125, NGC 5253 and Tol 89.
The faint broad \civ\ line requires sufficiently high S/N ($\ga$ 40) to be 
detected explaining the non-detection of this WC feature in previous observations. 
From the measurement of W-R emission lines (\niii+\Ciii, \heiia, and \civ), 
we conclude that all W-R regions contain a mixed  population of WNL, and early 
WC stars. The exception is the high-metallicity region NGC~3049 where late WC stars
prevail.

A spatial offset between the multiple peaks of the
nebular emission and the stellar light in He 2-10 and Tol 89 is observed. 
These nebular emission structures are likely due to the existence 
of bubbles and loops, owing to the injection of mechanical energy in the 
ISM through the W-R winds and/or supernovae. 
Due to age differences and likely smaller energy deposition the structures
around the W-R regions are possibly smaller than the ones predominantly 
energized by SNe.
The spatial distribution of W-R stars closely follows the stellar continuum 
with no significant distinction between WN and WC stars.

From the luminosity of the W-R signatures we have estimated the absolute number
of W-R stars of the different subtypes. The WC/WN number ratios  
have typical values between 0.2 -- 0.4, and show no clear trend with metallicity. 
For low-metallicity objects ($Z \sim 1/5$ \zsun), these values are larger than 
the observed WC/WN ratios in Local Group objects, but are compatible with 
expectations for star forming events with short duration
if stellar evolution models with high mass loss are used.
 
We derive ages for the starburst regions in the range of 3 to 6 Myr and confirm that
the burst duration must not exceed $\sim$ 2--4 Myr to account for 
the high population of W-R stars observed in starburst regions,
even if emission line stars similar to those observed in R136 and NGC~3603
are common in starbursts.
Within the uncertainties the majority of the observed quantities is 
reasonably reproduced by models with a Salpeter IMF. 
Although some W-R lines in few regions are stronger than predicted by the 
models no clear case requiring a significantly flatter IMF is found.
IMF slopes much steeper than Salpeter may, however, not be compatible 
with our data.

   \end{abstract}

%
%

\section{Introduction}
Starbursts play a major role in the global process of galaxy formation and 
evolution. However, despite two decades of intense research on starburst galaxies, 
our present knowledge of the intrinsic starburst properties is still rather 
poor and fundamental questions remain to be answered. 
For example, what is the duration of starbursts ?
Which type of stellar population is formed during these events ?
Is the proportion of massive stars higher in starbursts than in more 
``quiet'' star-forming regions?
Observationaly, answering these questions is not an easy task. 
In this respect the so-called ``Wolf-Rayet (W-R) galaxies'' 
may be the ideal laboratories since these objects harbour the most massive 
stars known,
O stars and their descendents W-R stars, which allow to probe the upper 
part of the initial mass function (IMF) and the youngest stellar 
populations.
    
W-R galaxies are usually defined as ``those galaxies in whose
integrated spectra a broad emission feature at \heiia\ attributed to
W-R stars has been detected'' \cite{C91}. In practice, the detection
of a broad emission feature at $\lambda 4650 - 4690$ (the so-called ``W-R bump'',
which possibly includes additional emission lines, cf.\ below) attributed to W-R stars 
is often simply used.

Whereas the initial compilation of Conti~\cite*{C91} includes 37 galaxies, 
to date at least 100 such objects are known (see the compilation of Schaerer 
\& Contini, 1998). 
W-R galaxies are found  among emission-line galaxies.
However, the class of W-R galaxies encompasses a wide range of 
properties and galaxy types. While most of them fall in the category of
\hii\ galaxies, broad W-R emission lines have also been detected in more
``exotic'' objects like
Luminous Infrared Galaxies (Armus \etal, 1988),
Seyfert 2 galaxies (e.g.\ Heckman \etal, 1997)
and giant cD galaxies, located in the centre of galaxy clusters with 
strong cooling flows \cite{A95}.

The mere presence of W-R stars in these objects indicates recent ($\la 10$ Myr)
star formation and the existence of massive stars ($M_{\rm initial} \ga 25$ 
\msun, cf.\ Maeder \& Conti, 1994); this provides already interesting information
about star formation in these objects.
Quantitative analysis of the \heiia\ emission or the W-R bump
show that star formation
must have occurred over short timescales compared to the lifetime of massive
stars \cite{KS81,ARNAULTetal89,VC92,M95}, 
and have allowed to derive the absolute number of W-R stars present in these 
regions ($N_{\rm W-R} \sim 100 - 10^5$, e.g.\ Vacca \& Conti 1992, hereafter
VC92).

This blue W-R bump is often blended with nearby nebular emission lines
of He, Fe, or Ar, and can show several broad stellar emission components
(\niii, \Ciii, \heiia) difficult to separate in most low or medium-resolution 
spectra. Their origin can a priori be due to W-R stars of WN and/or WC subtypes,
which has been a question of some earlier dispute (cf.\ Osterbrock \& Cohen, 1982; 
Sargent \& Filippenko, 1991; with Kunth \& Schild, 1986; Conti, 1991). 

The strongest emission line of WC stars is \civ\ which is very weak in WN stars.
This ``red W-R bump'' has quite rarely been observed so far. Although Conti \cite*{C91} 
claimed the absence of convincing evidence for carbon features in W-R galaxies
(cf.\ also Kunth \& Schild, 1986) there is no doubt about their existence anymore
(e.g.\ NGC 4861: Dinerstein \& Shields, 1986; NGC 4214: Sargent \& Fillipenko, 1991, 
Kobulniky \& Skillman, 1996; 
NGC 2363: Gonzalez-Delgado \etal, 1994; Mrk 996: Thuan, Izotov \& Lipovetsky, 1996; 
Mrk 475, Mrk 1450: Izotov \etal, 1994;
NGC 7714: Garcia-Vargas \etal, 1997; I Zw 18: Izotov \etal, 1997, Legrand \etal, 
1997b). 
Where the data is available, \civ\ is generally weaker than \heiia.
In our recent compilation \cite{SC98} we find $\sim$ 15 
objects with a fairly well established detection of broad \civ.

The presence of WC stars, although possibly less numerous than WN stars, 
is indeed expected both from observations of W-R populations in the Local Group 
(e.g.\ Massey \& Johnson, 1998) and from stellar evolution models
\cite{MM94}. The predictions are that, depending on the 
evolutionary model and  metallicity, $0 - 55$ \% of young starburst should 
show an important WC population (Meynet, 1995; Schaerer \& Vacca, 1998; hereafter SV98).
Most surprisingly the homogenenuous, fairly high signal-to-noise spectra from the sample 
of VC92 containing 12 regions with \heiia\ detections (and 10 upper limits) 
revealed only one \civ\ detection (in He 2-10 A). 
If true, this would certainly contradict the predictions from synthesis
models (Meynet, 1995; SV98) using high mass loss stellar evolution models, 
which otherwise compare well to observations in the Local Group \cite{MM94}.

To verify if this apparent discrepancy really holds we have initiated a 
search for WC stars in well-known W-R galaxies. We observed five objects (NGC~3049, 
He 2-10, NGC~3125, NGC~5253, Tol~89), three of them are in common with VC92. 
A first account of our observations of NGC~5253 was given in Schaerer \etal\
\cite*{SCHAERERetal97}. 

In the present paper we report the successful discovery of both WN and WC 
stars in all of the objects, which supports our initial hypothesis of an 
observational bias (i.e.\ a too low signal-to-noise ratio around 5800 \AA) against 
their detection in the VC92 sample, and (at least partly) relieves the 
aforementioned discrepancy.
We then use both the WN and WC signatures to constrain the burst properties
and stellar evolution models.

The paper is structured as follows. The observations and reductions are
described in Sect.\ 2. A spatial analysis of the emission lines is 
shown in Sect.\ 3. We describe the properties of the W-R regions
and their massive star content in Sect.\ 4. 
Constraints on the evolutionary tracks and the properties of the starburst
regions (age, burst duration, IMF) are derived in Sect.\ 5 from 
a comparison with evolutionary synthesis models. Finally, our main results
are summarised and discussed in Sect.\ 6.

\begin{table*}
\caption[]{Global properties and observing log of galaxies}
{\scriptsize
\begin{tabular}{lcclccrrrcrr}
\hline
\hline
Galaxy & \multicolumn{2}{c}{Coordinates (J2000)}  & Type & $m_{\rm B}$ & 
E(B-V) & $V_{\odot}$ & $D$ & $M_{\rm abs}$ & Exposure time & P.A. & Scale \\
 & $\alpha$ & $\delta$ & & [mag] & [mag] & [\kms] & [Mpc] & [mag] & [min] & [\degr] & [pc/\arcsec] \\
\hline
He 2-10  & 08$^{\rm h}$ 36$^{\rm m}$ 16\fs0 & $-$26\degr 24\arcmin 40\arcsec & I0 Pec 
& 12.5   &  0.21 & 873$\pm$10 & 8.8 & -17.3 & 80 (4$\times$20) & 94 & 43\\
NGC 3049 & 09$^{\rm h}$ 54$^{\rm m}$ 49\fs9 & $+$09\degr 16\arcmin 19\arcsec & SB(rs)ab 
& 13.0   & 0.04 & 1494$\pm$4 & 18.3 & -18.3 & 80 (4$\times$20) & 35 & 89\\
NGC 3125 & 10$^{\rm h}$ 06$^{\rm m}$ 34\fs4 & $-$29\degr 56\arcmin 10\arcsec & E?    
& 13.3   & 0.25 & 1080$\pm$47   & 11.5 & -17.0 & 100 (5$\times$20) & 123 & 56\\
NGC 5253 & 13$^{\rm h}$ 39$^{\rm m}$ 55\fs8 & $-$31\degr 38\arcmin 41\arcsec & E/S0?   
& 11.1   & 0.05 &  404$\pm$4   & 4.1 & -17.2 & 90 (4$\times$20 + 10) & 24 & 19\\
%
Tol 89   & 14$^{\rm h}$ 01$^{\rm m}$ 25\fs4 & $-$33\degr 04\arcmin 28\arcsec &  
G\hii\ in SBdm & 16.0   & 0.23 &  1226$\pm$11   & 14.7 & -14.8 & 120 (6$\times$20) & 39 & 71\\
\hline
\hline
\noalign{\smallskip}
\noalign{\small Global parameters come from RC3 except for the heliocentric radial 
velocity ($V_{\odot}$) of He~2-10 which comes from Kobulnicky \etal\ \cite*{KOBULNICKYetal95}. 
The Galactic foreground extinction E(B-V) comes from Burstein \& 
Heiles~\cite*{BH84}. 
The distance ($D$) is derived from the Galactic Standard 
of Rest (GSR) velocity, using a Hubble constant $H_0 = 75$ \kmsmpc, 
except for the distance of NGC~5253 taken from Saha \etal\ (1995).
The absolute blue magnitude ($M_{\rm abs}$) is derived from the apparent magnitude 
($m_{\rm B}$) and the adopted distance ($D$). 
The position angle (P.A.) of the slit is measured from North to East. The linear scale (in pc/\arcsec) 
is computed using the adopted distance adopted distance ($D$)}
\end{tabular}
}
\label{GLOBAL}
\end{table*}


\section{Observations and data reduction}

Long-slit spectra of galaxies were obtained on the nights of 1995 April 24 -- 
26 at the ESO 2.2m telescope. The data were acquired with the EFOSC2 
spectrograph and a 1024 $\times$ 1024 Thomson CCD with a pixel size of 
0.34\arcsec. We used grism 4 which gives a spectral coverage of 4400 -- 6500 
\AA\ with a resolution of $\sim$ 5 \AA. 
During the nights, we also 
observed the spectrophotometric standard stars HD~84937 and Kopff~27 in order 
to flux calibrate the spectra of the galaxies. Note that only the first night 
was photometric. Spectra of He-Ar calibration 
lamp were obtained immediately before and after the galaxy integrations in 
order to accurately calibrate the wavelength scale.

The slit was oriented in order to cover regions with previous 
detections of W-R stars and others bright optical regions in galaxies. 
The slit width was 1.6\arcsec\ for the galaxy observations and 5\arcsec\ 
for the standard stars. The total integration time for each galaxy is given 
in Table~\ref{GLOBAL}. We took multiple exposures of 1200 s in duration, 
short enough to avoid saturation of the bright nebular lines (\hbeta\ 
and \oiii) and to recognize cosmic ray impacts. The seeing was relatively 
stable during the observations with a mean spatial resolution of about 
1\arcsec, as measured by the La Silla seeing monitor. 
The spectra were acquired at low airmass ($\leq$ 1.1, except for 
NGC~3049 observed with an airmass $\sim$ 1.3) and no correction for the loss 
of blue light due to atmospheric dispersion was made.

The spectra were reduced according to  standard reduction 
procedures using the MIDAS package LONG. 
A bias level was subtracted from each frame using the overscan region of the 
CCD chip. A combination of flat field images from the twilight sky and 
a continuum 
quartz lamp were divided into each frame to correct for pixel to pixel 
variations along spatial and dispersion axes respectively. 
After wavelength calibration, transformation to equal wavelength intervals 
along the dispersion axis and flux calibration using standard stars 
observations, spectra from the same galaxy were combined to average frames 
and reject cosmic ray events. 
Care was taken to discriminate 
between cosmic rays and the high peaks of strong emission lines which might 
also be confused with cosmic ray events. Tracking with the autoguider was 
sufficiently good that no shifting or registration of the frames was required 
prior to the combine procedure. 
Emission-lines from the night sky were then subtracted using a linear fit to 
at least 50 ``sky'' pixels on either side of the galaxy. 

\section{Detailed spatial analysis}
\label{SPATANAL}

We first use our long-slit spectroscopic observations to compare, 
for each galaxy, 
the spatial distributions of nebular and W-R emission-line intensities 
and the stellar continuum emission.
For this purpose, we extracted one-dimensional spectra along the slit 
by adding  3 pixels to match the seeing of our
observations ($\sim$ 1\arcsec). We choose one pixel of overlap between two 
adjacent spectra to maximize signal-to-noise ratio while still maintaining a 
good spatial resolution. 
Depending on the spatial extent we performed between 80 and 180 extractions per galaxy, 
such that the brightest emission-lines 
(\hbeta\ and \oiiib) were strong enough 
to be measured with reliability.

We then measured, for each individual spectrum, the parameters 
(central wavelength, FWHM, flux and equivalent width) of the brightest 
nebular emission lines (\heia, \hbeta, \oiiia, \oiiib, \heic\ and \oia), 
the broad emission lines due to W-R stars 
(blue W-R bump: blend of \niii\ and \Ciii, \heiia; red W-R bump: \civ) 
together with the continuum flux under each emission line. 

\subsection{Variations of emission line and continuum intensities}


\begin{figure}[t]
  \resizebox{\hsize}{!}{\includegraphics{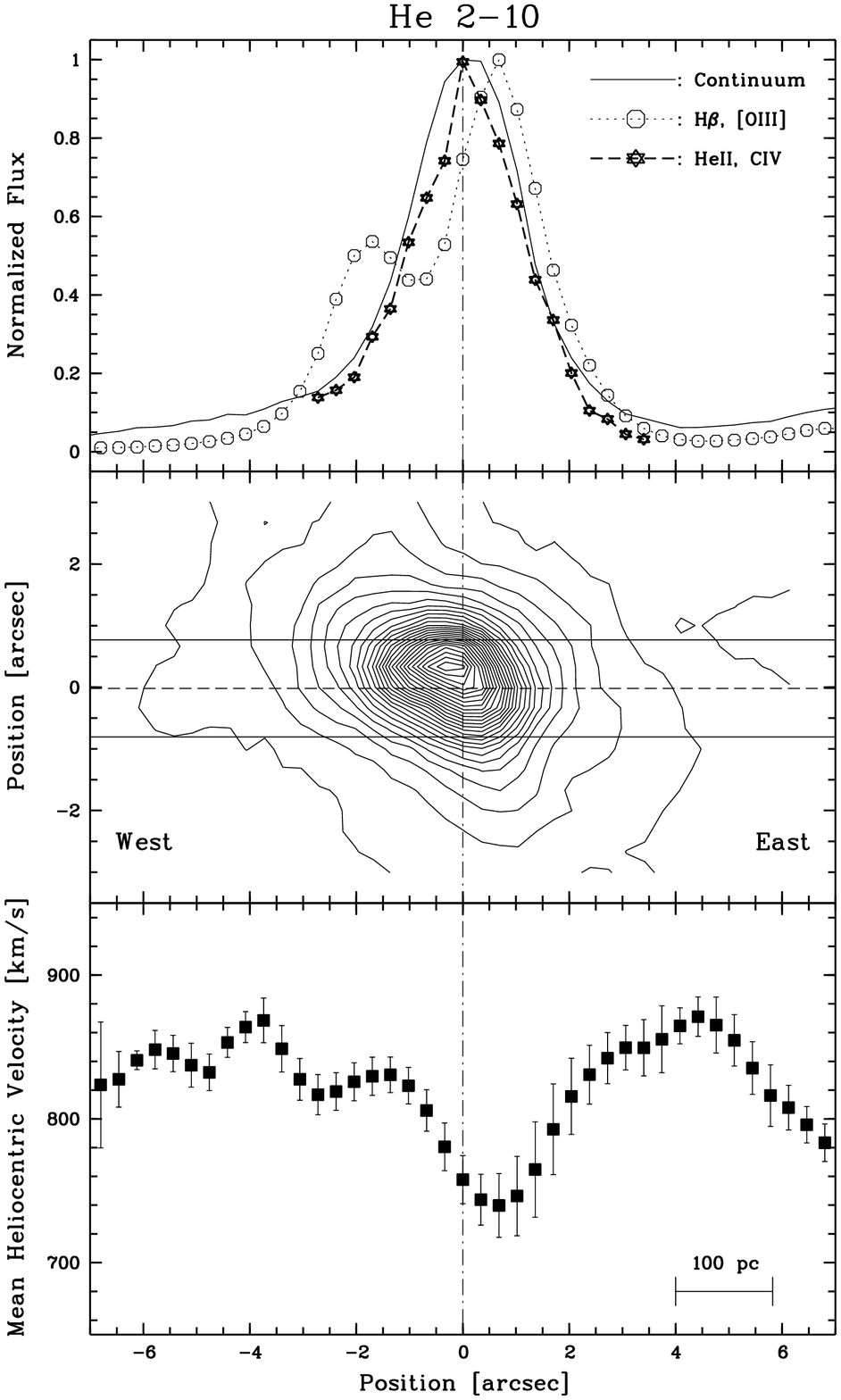}}
  \caption{He~2-10.
{\it Top}: Spatial distribution of the stellar continuum (solid line), of 
the brightest nebular emission lines (\hbeta, \oiiib, dotted line) and of 
the broad emission lines (\heiia\ and \civ, dashed line) due to W-R stars. 
Capital letters, if present, indicate the position of different WR regions
(see Table \ref{DATA}).
{\it Middle}: Isophotal contour map of the galaxy showing the position of 
the slit used for the spectroscopic observations. 
{\it Bottom}: Velocity curve of the ionized gas along the slit. 
The position of peak intensity from W-R stars is indicated by the vertical 
line. Position along the slit is given in arcsec. The scale corresponding 
to 100 pc is given on the bottom right corner of the figure
  }
  \label{HE0210_SPAT}
\end{figure}

\begin{figure}[t]
  \resizebox{\hsize}{!}{\includegraphics{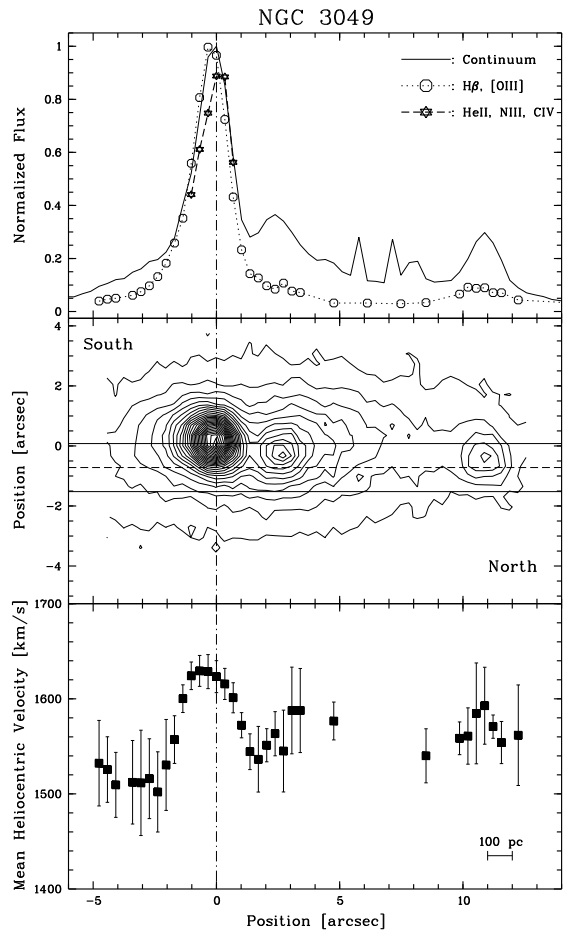}}
  \caption{NGC~3049. See Fig.~\ref{HE0210_SPAT} for the legend}
  \label{N3049_SPAT}
\end{figure}

\begin{figure}[t]
  \resizebox{\hsize}{!}{\includegraphics{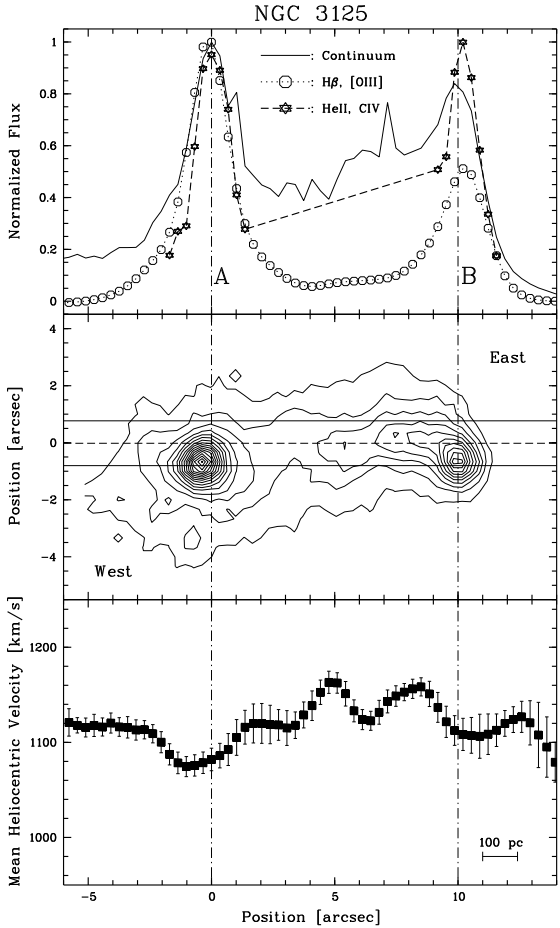}}
  \caption{NGC~3125. See Fig.~\ref{HE0210_SPAT} for the legend}
  \label{N3125_SPAT}
\end{figure}

\begin{figure}[t]
  \resizebox{\hsize}{!}{\includegraphics{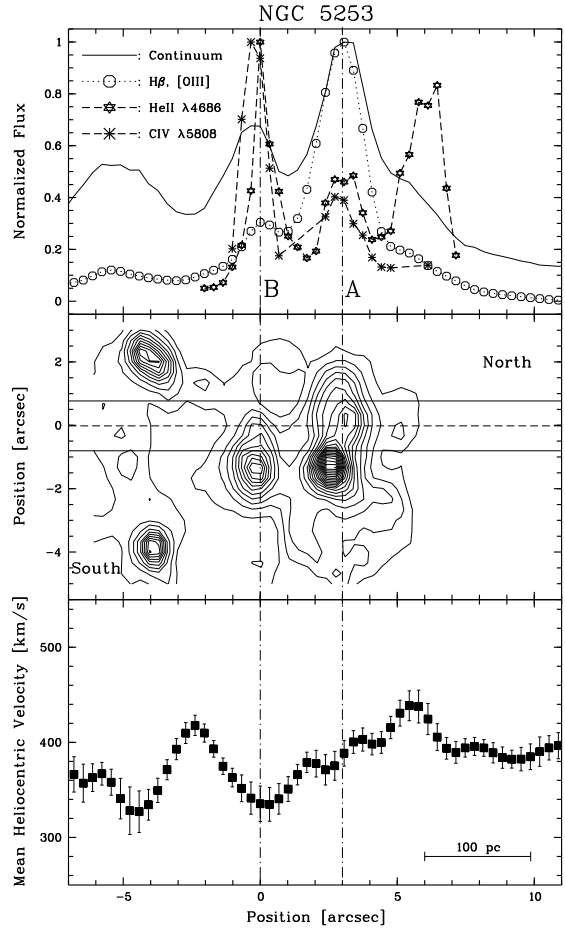}}
  \caption{NGC~5253. See Fig.~\ref{HE0210_SPAT} for the legend}
  \label{N5253_SPAT}
\end{figure}

\begin{figure}[t]
  \resizebox{\hsize}{!}{\includegraphics{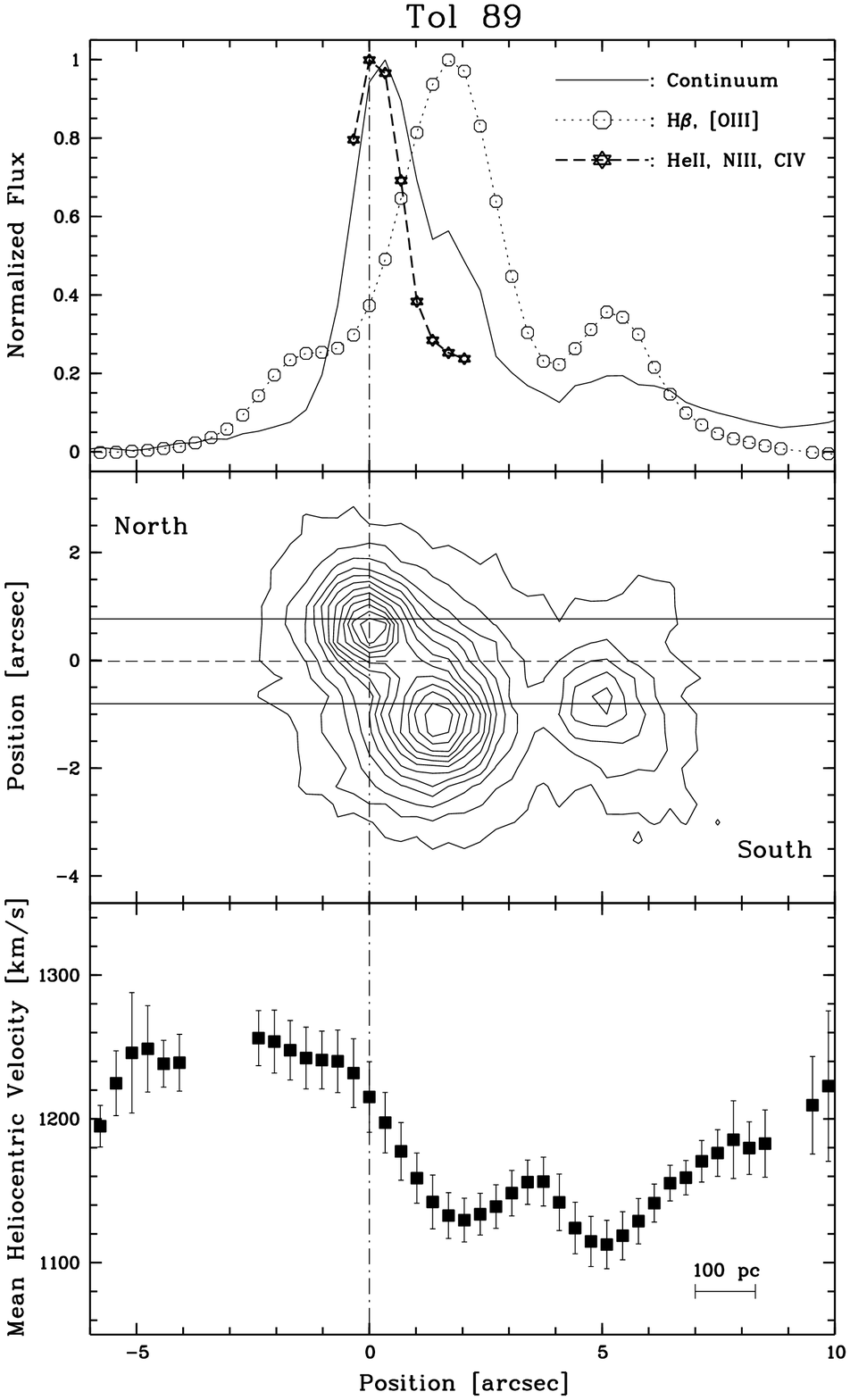}}
  \caption{Tol~89. See Fig.~\ref{HE0210_SPAT} for the legend}
  \label{TOL89_SPAT}
\end{figure}


The results of the spatial analysis are displayed in Figs.~\ref{HE0210_SPAT} 
to \ref{TOL89_SPAT}.
We do not observe any significant variation in the spatial distribution 
of the bright nebular emission lines. We therefore plot (see top panel) 
the mean nebular intensity distribution corresponding to the averaged flux of 
the brightest lines \hbeta\ and \oiiib. 
The same applies to the continuum emission; no significant variation is seen 
in the intensity distribution from 4400 \AA\ to 6500 \AA, at least for the 
small portion of galaxy shown in the figures.
A small shift between the position of peak intensity of the blue and 
red W-R bumps is observed in nearly all the galaxies. This difference is 
however too small ($<$ 1\arcsec) to be significant, given the spatial resolution 
of our observations. We thus plotted the mean intensity distribution of 
W-R emission lines by averaging the intensities of \heiia, \civ\ and 
\niii+\Ciii\ (for NGC 3049 and Tol 89 only). 

In the middle panel of each figure, 
we include an isophotal map of the galaxy produced from a CCD image taken 
without filter
just before the spectroscopic observation. The position and direction of 
the slit are indicated. This image is very useful to identify the 
peak intensities of the continuum and emission lines. 
In the bottom panel, we show a velocity curve derived from the measured 
central wavelength of 
nebular emission lines along the slit. The mean radial heliocentric velocity 
has been derived using the brightest nebular lines, i.e. 
\hbeta, \oiiia, \oiiib\ and \heic, since no systematic variation in the 
velocity curves has been observed for the various emission lines.

One of the most striking features which appear in Figs.~\ref{HE0210_SPAT} and 
\ref{TOL89_SPAT} is the difference between the distributions of stellar 
continuum and nebular lines in He~2-10 and Tol~89. The stellar continuum 
shows only one maximum whereas the distribution of nebular lines intensity is 
mostly double-peaked. Moreover, the maximum intensity of nebular lines is 
shifted relative to the maximum intensity of stellar continuum, both for 
He~2-10 ($\sim$ 1\arcsec\ to the East) and Tol~89 ($\sim$ 2\arcsec\ to the 
South-West). 
In the other galaxies, i.e. NGC~3049, NGC~3125 and NGC~5253, the peaks of 
the nebular lines and stellar continuum intensity distribution are coincident. 
  
In all the galaxies, the peak intensity of the broad emission lines \heiia\ 
and \civ\ coincide with the peak intensity of the stellar continuum, even 
in He~2-10 and Tol~89 where there is a shift between the distributions 
of stellar continuum and nebular lines. These results confirm the stellar 
origin of the broad W-R emission lines.
Given the good signal-to-noise ratio of our observations, it has been possible 
to detect the W-R lines over a large extension around the maximum intensity. The 
size of the ``W-R regions'' ranges from 2\arcsec\ -- 3\arcsec\ for Tol~89, 
NGC~3049, NGC~3125 and NGC~5253 to 6\arcsec\ for He~2-10. The W-R lines are 
detected in two distinct regions in NGC~3125 (separation $\sim$ 10\arcsec) 
and NGC~5253 (separation $\sim$ 3\arcsec). See Table~\ref{GLOBAL} for the 
corresponding linear scale (in pc/\arcsec) of the galaxies. 

\subsection{Spatial distribution of Wolf-Rayet stars}
\label{W-RDISTRI}

In this Section we shall discuss the spatial distribution
of the W-R signatures in the individual objects. To this end, we use 
(when available) high resolution images from the literature to identify the 
W-R regions. 

\subsubsection*{He~2-10}
Three starburst regions can be distinguished on the 
optical images of He~2-10 \cite{HS84,CORBINetal93}. Following the nomenclature 
of Corbin \etal\ \cite*{CORBINetal93}, the most prominent is region $A$ at 
the center of the galaxy. 
The second starburst region $B$ is located at $\sim$ 8.5\arcsec\ East of $A$. 
It is smaller and fainter in optical images and is seen at an even lower 
level in \halpha. The third region called $C$ \cite{SAUVAGEetal97}, 
located $\sim$ 2.5\arcsec\ West of region $A$, is not prominent in optical 
images but is more noticeable in \halpha\ and \oiii\ images \cite{CORBINetal93}.
Regions A, B, and C are resolved into several knots on {\it HST} UV images 
\cite{CV94}.
The presence of three starburst regions in the central part of He~2-10 
is consistent with the spatial distribution of W-R emission lines along 
the slit ($\sim$ 4\arcsec\ long) which shows a maximum in $A$ but 
extends also to the region $C$, the secondary peak of nebular emission seen in 
Fig.\ref{HE0210_SPAT}. 
Knot $B$ was not included in our slit. 
Our long-slit spectrum of He~2-10 shows W-R features in regions A 
and C. However, the spectral signatures in region $C$ are too faint to be 
considered separately from region $A$ in the following analysis (see 
Sect.~\ref{W-RREGIONS}).

\subsubsection*{NGC~3049, NGC~3125} 
In NGC~3049 and NGC~3125, the W-R regions coincide 
with the peaks of the continuum emission and nebular lines distributions.
NGC~3049 is a barred spiral galaxy belonging to the Virgo cluster. 
This low-mass galaxy has a small bulge and a thin bar of constant 
surface brightness surrounded by an inner ring \cite{CONTINIetal97}. 
The optical appearance of NGC~3125 is an amorphous elliptical 
shape with a bright central starburst region dominated by two bright knots 
(regions $A$ and $B$) apparently connected by a bridge of fainter intensity. 

\subsubsection*{NGC~5253}
Our observations of NGC~5253 have already been presented in Schaerer et al. (1997).
The spatial distribution of W-R stars in this object is complex 
(see Fig.~\ref{N5253_SPAT}).
There are two W-R regions, labeled as $A$ and $B$, located at the two peaks 
intensity of the continuum and nebular lines distributions.
The brightest W-R region ($B$), however, corresponds to the faintest continuum and 
emission lines region. For this galaxy we made a 
distinction between the distribution of the \heiia\ and \civ\ lines. There is 
indeed a third bright ``\heiia\ region'' (region $H$ in Schaerer \etal, 1997) 
located 6\arcsec\ North-East of the brightest W-R region $B$. The spectrum 
of this region show a narrower \heiia\ line than in others regions $A$ and 
$B$, probably of nebular origin \cite{SCHAERERetal97}. In fact, the 
core of the galaxy ($\sim$ 20\arcsec\ of diameter) host a dozen blue 
stellar clusters as well as diffusely distributed massive stars 
(Meurer \etal\ 1995 ;cf.\ also Calzetti \etal, 1997 for a recent study
of NGC~5253). 

\subsubsection*{Tol~89}
Tol~89 is a giant \hii\ region located in one of the spiral arms of the 
barred spiral galaxy NGC~5398, about 34\arcsec\ South-West of the nucleus.  
This object shows the largest shift between the spatial distribution of the
continuum and nebular lines.
The spatial distribution of the W-R lines is narrower than that of the nebular
lines. W-R lines are found at the two regions where the continuum peaks.

\section{The Wolf-Rayet regions}
\label{W-RREGIONS}


\begin{figure}[t]
  \resizebox{\hsize}{!}{\includegraphics{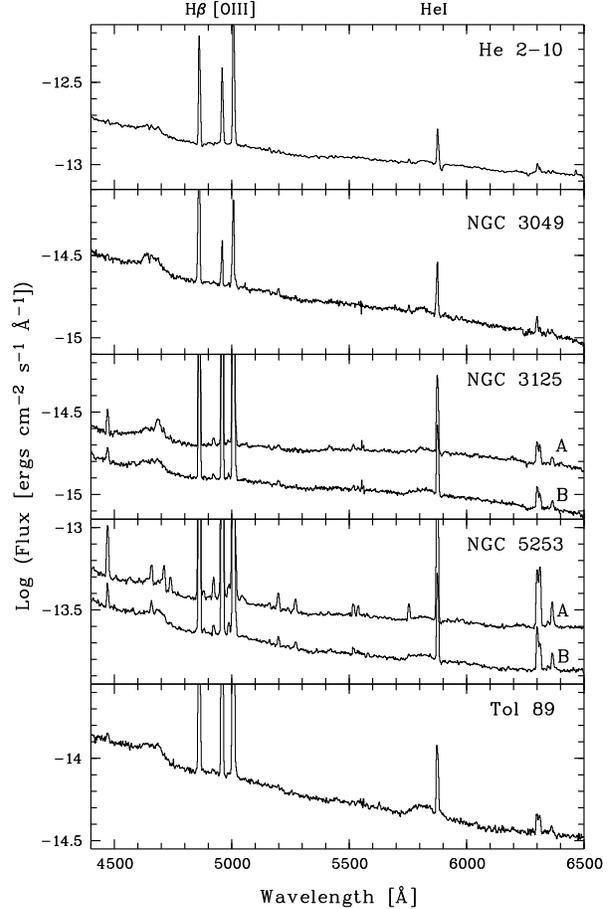}}
  \caption{Optical spectra of W-R regions at the rest frame wavelength of galaxies. 
Note the blue continuum and the bright nebular emission lines (\hbeta, \oiii, and 
\heia) typical of starburst regions. Normalized spectra showing the region around 
the blue (around 4700 \AA) and red (around 5800 \AA) W-R bumps are given in 
Figs.~\ref{SPEC4700} and \ref{SPEC5800} respectively. The spectra are not corrected 
for extinction
  }
  \label{SPECALL}
\end{figure}

\begin{figure}[t]
  \resizebox{\hsize}{!}{\includegraphics{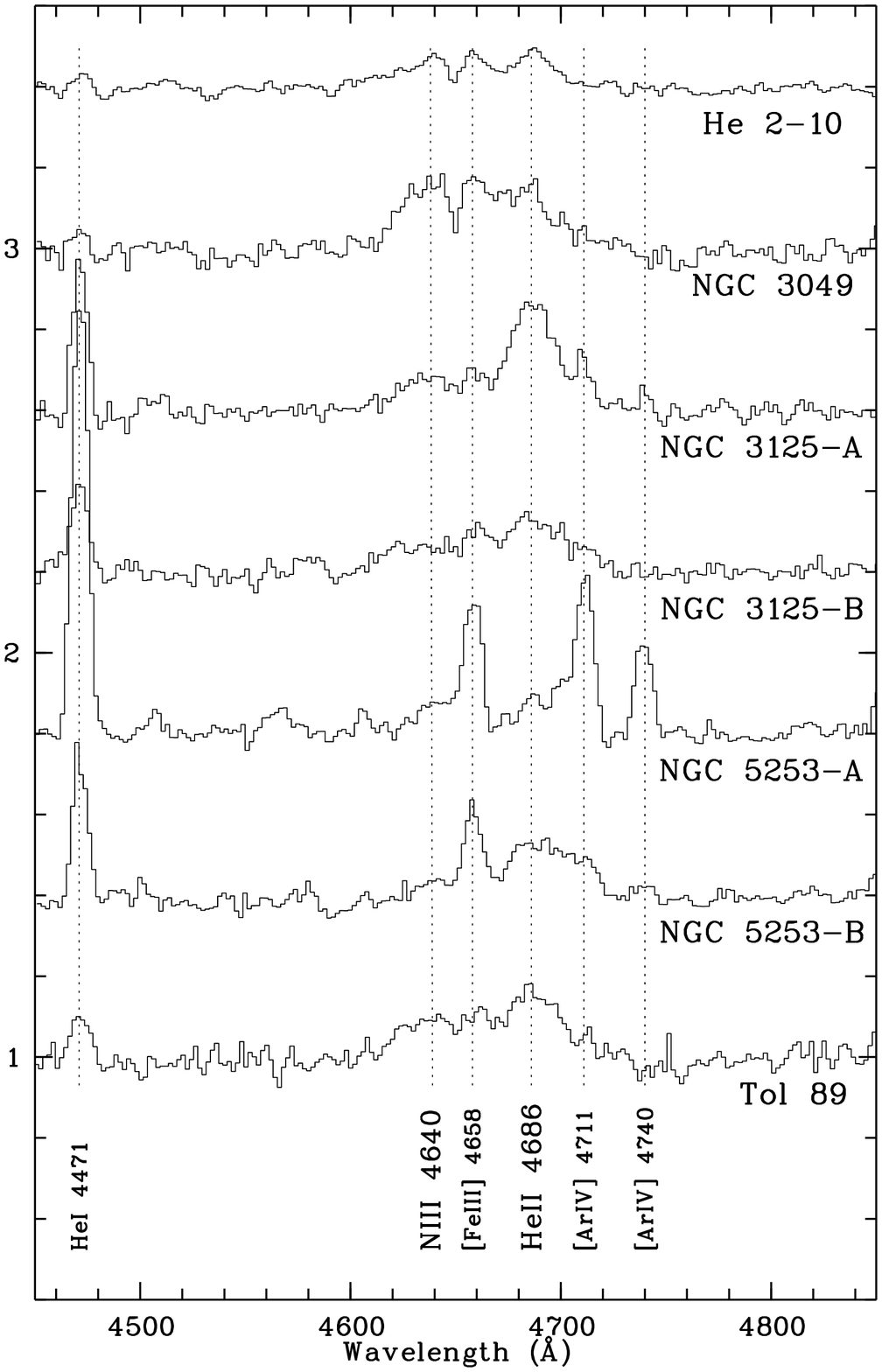}}
  \caption{Enlargement of the blue W-R bump in the optical spectrum of galaxies. 
Spectra are normalized to the continuum and an offset of 0.4 has been applied 
between each spectrum. 
The broad \heiia\ line from WNL stars is clearly detected. The broad line 
around 4645 \AA\ is not well identified and could be a blend of \niii\ and \Ciii\ 
lines due to WN and WC stars. The spectra of NGC~3125-A and NGC~5253-A exhibit 
the typical high-excitation nebular emission lines of \ariva\ and \arivb. Their 
presence might imply a nebular contribution to the \heiia\ line
  }
  \label{SPEC4700}
\end{figure}

\begin{figure}[t]
  \resizebox{\hsize}{!}{\includegraphics{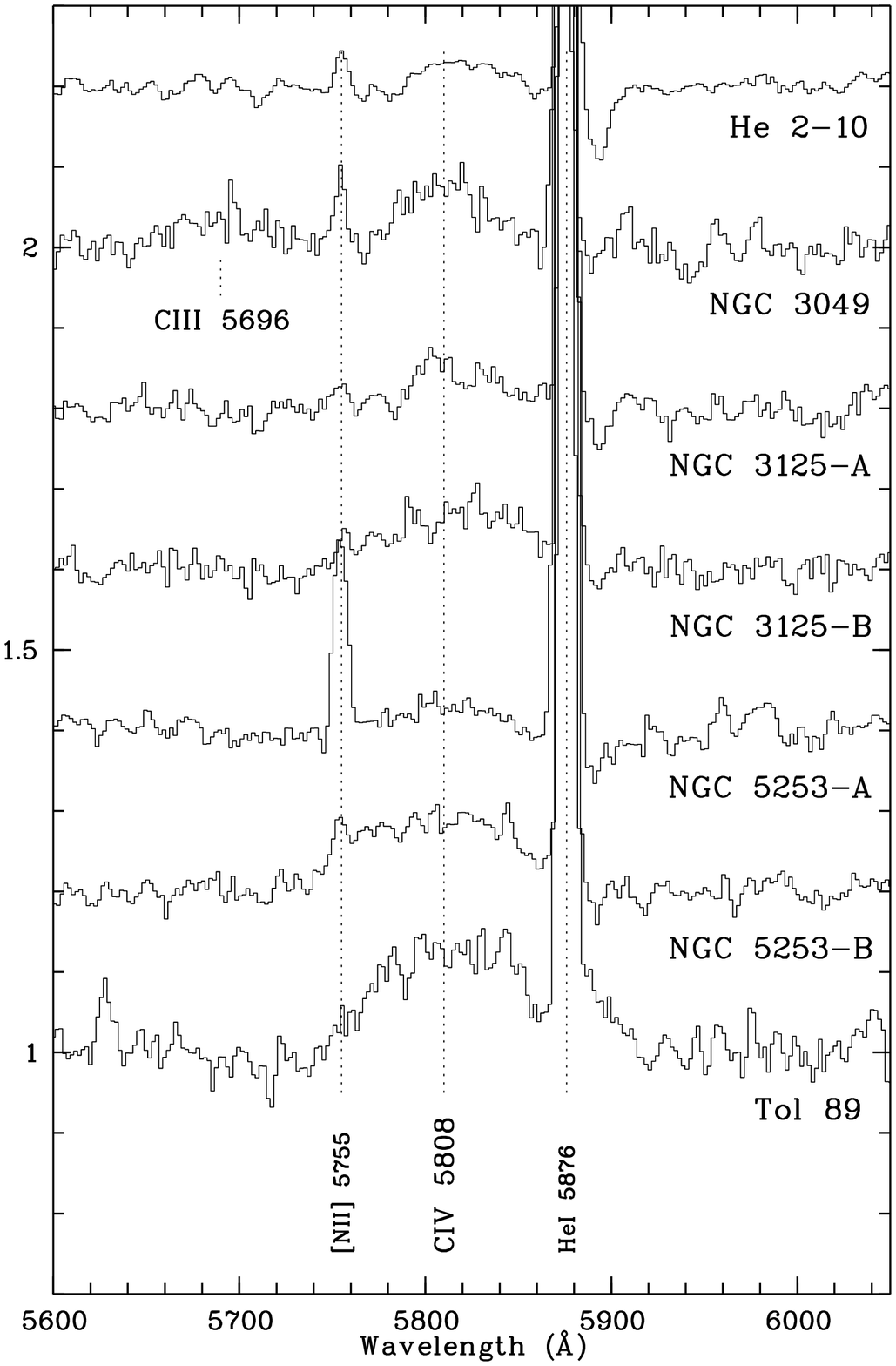}}
  \caption{Enlargement of the red W-R bump in the optical spectrum of galaxies. 
Spectra are normalized to the continuum and an offset of 0.2 has been applied 
between each spectrum. The broad \civ\ line from WC stars is clearly detected 
in all spectra. Except for He~2-10, this represents the first detection of WC 
stars in these galaxies. Note also that a broad emission line of \ciii\ 
is also reported at $\sim 2 \sigma$ level in the spectrum of NGC~3049. This 
indicates the predominance of late-type WC stars in this high-metallicity galaxy
  }
  \label{SPEC5800}
\end{figure}

In this section, we present a spectroscopic analysis of the individual 
regions where spectral signatures from W-R stars have been detected. 
One-dimensional spectra were extracted corresponding to the different
regions by adding several columns along the slit. The size of 
the extraction window has been chosen for each region in order to obtain 
the best compromise between a high signal-to-noise ratio in the continuum 
(around 4700 \AA\ and 5800 \AA) and a good enough spatial resolution. 
We extracted one-dimensional spectra centered on the maximum intensity of 
W-R emission lines with an aperture of $\sim$ 1.7\arcsec\ 
for NGC~5253 (regions $A$ and $B$) and Tol~89, $\sim$ 2.0\arcsec\ for 
NGC~3049, $\sim$ 2.7\arcsec\ for He~2-10 and NGC~3125 (regions $A$ and $B$). 
We thus obtained spectra for a total of seven
regions, one per galaxy except for NGC~3125 and NGC~5253 in which two W-R 
regions have been identified. The full wavelength range ($4400 - 6500$ \AA, 
corrected for the Doppler effect) of the flux calibrated spectra (not corrected 
for reddening) of W-R regions are shown in Fig.~\ref{SPECALL}. 
Normalized spectra showing the region around the blue and red W-R bumps are 
given in Figs.~\ref{SPEC4700} and \ref{SPEC5800} respectively. 

Given the complex structure and multiple blends around the blue W-R bump
and the medium resolution of our spectra we adopt the following 
procedure to measure accurately the individual emission lines.
We use a procedure 
of multi-gaussians fitting where the number of gaussians is the only 
fixed parameter (central wavelength, intensity and FWHM are free to vary). 
The normalized spectra were obtained by dividing the spectra by a 
low-order polynomial fit on the line-free region of the continuum.
In Table~\ref{DATA} we report the measurements performed on the 
normalized spectra. The uncertainties are discussed next.

\subsection{Uncertainties}
\label{UNCERTAINTIES}

There are three major sources of uncertainties in the measured fluxes and 
equivalent widths of emission lines. The first one is inherent to the 
measurement itself and depends mainly on the signal-to-noise ratio of the spectra.
These errors were computed directly when using the IRAF task SPLOT to derive 
the parameters of emission lines. The second source of errors comes from 
the extraction of an individual spectrum corresponding to a region in a galaxy.
Indeed, for some galaxies like He~2-10 and Tol~89, we observe a shift 
between the maximum emission of the stellar continuum and of the nebular 
lines (see Sect.~\ref{SPATANAL}). Thus, the measured line flux and 
equivalent width of the lines
is sensitive to the size of the extraction window used to obtain 
one-dimensional spectra of the individual regions. To estimate 
this error, we measured the lines in  extracted spectra  from effective
 apertures sizes ranging from 1\arcsec\ to 4\arcsec. 
A third source of uncertainty exists for lines from the blue W-R bump.
Given the important slope of spectra and the complex structure in this
region the line measurement also depends on the determination of the
continuum level (i.e.\ the normalization).
Therefore measurements of emission lines in the W-R bumps were performed 
before and after normalization. 

The three sources of errors were propagated quadratically to obtain the 
final uncertainties reported in Table~\ref{DATA}
for the dereddened line fluxes and for the equivalent widths.
For the bright nebular emission lines (like \hbeta, \oiii\ and \heic), 
the relative uncertainty is generally smaller than 
10\%. However, it can reach a 
maximum of $\sim$ 20\% for the equivalent width 
and of $\sim$ 30\% for the relative flux of faintest broad emission lines from 
W-R stars. 
We find that generally errors due to the 
measurement itself (which is $\propto$ S/N) and to the extraction window are 
comparable. For emission lines from the blue W-R bump, the uncertainties due 
to the normalization are not negligible and can even be comparable in some cases. 
We note however some exceptions. For He~2-10 and Tol~89, which 
show a discrepancy between the nebular and the continuum emission, the uncertainty 
on the equivalent width of nebular lines (and especially for \hbeta) is dominated 
by the error due to the size of the extraction window. 

We checked the absolute flux calibration (\hbeta\ flux), the relative line 
flux of \heia, \oiiia\ and \heic\ lines, and the equivalent width of \hbeta\ 
by comparing our measurements with previous observations 
reported in the literature: 
He~2-10 \cite{VC92}, 
NGC~3049 \cite{KS86,VC92,CONTINIetal97}, 
NGC 3125 \cite{KS81,KS83,KJ85,VC92}, 
NGC~5253 \cite{CAMPBELLetal86,W-R89,TERLEVICHetal91}, 
and Tol~89 \cite{DURRETetal85,TERLEVICHetal91}. 
A good agreement was generally found for the absolute calibration and 
the relative fluxes of bright nebular lines, with relative errors $\la$ 20\%. 

Larger differences can be found for the broad W-R lines.
Comparing our data  with VC92, our measurements of \heiia\ 
(relative flux to \hbeta\ and equivalent width) are found to be systematically 
larger by a factor of $\sim$ 3 in NGC~3049, and $\sim$ 1.5 in He~2-10 and 
NGC~3125, although a good agreement is found for the nebular lines.
Given the systematic differences, it is unlikely that different slit positions
are responsible for this effect.
To understand the origin of this discrepancy, we compared our spectra
to those published in VC92 and kindly provided by William Vacca.
The comparison shows that the differences can mostly be understood
in terms of differing signal-to-noise ratios. 
Indeed, the measurement of the faint, broad W-R emission lines is
strongly dependent on the signal-to-noise ratio, especially to fit the continuum 
level in the deblending process used to separate the W-R bump into the 
individual components. 
Given the improved quality of our spectra (S/N $\sim$ 50--70 around 4700 \AA\
compared to S/N $\la 40$ for VC92), our measurements should be more reliable.
In particular the improved S/N also allows the detection of the weaker
\niii+\Ciii\ blend, barely detectable in the data of VC92.

Within the uncertainties, our new measurements of W-R emission lines in 
NGC~5253 are consistent with those given in Schaerer \etal\ \cite*{SCHAERERetal97},
which were obtained from a different reduction/analysis. 
 
\subsection{Reddening}
\label{REDDENING}

Due to the limited spectral range, our observations do not include the 
\halpha\ emission line commonly used to determine the internal reddening 
parameter \cbeta. 
We therefore adopt values taken from the recent literature.
We wish to remind that the determination of the amount of reddening is 
subject to two complications often not accounted for: 
1) the value of the extinction coefficient can be wavelength dependent,
and
2) variations of the extinction on small spatial scales can occur.
The former is well illustrated by the case of He~2-10, where
\cbeta\ varies from $\sim$ 0.4 in the optical 
\cite{ST92,VC92} to $\sim$ 4.5 in the infrared \cite{AR84,KAWARAetal89} and 
can even reach $\sim$ 14 in the millimeter range \cite{KOBULNICKYetal95}.
NGC~5253 is a good example showing large spatial variations of reddening
\cite{W-R89,CALZETTIetal97}.
The adopted extinction coefficient is taken from
regions corresponding as closely as possible to the ones studied here
and derived from optical spectra, i.e.\ from the Balmer decrement.
Our adopted reddening values
are given in Table~\ref{DATA}. 

For He~2-10 and NGC~3125, we adopt the internal extinction coefficients 
given by VC92, with different values for regions 
$A$ (\cbeta\ $\sim$ 0.40) and $B$ (\cbeta\ $\sim$ 0.64) of NGC~3125. 
For NGC~3049, we choose to use the value \cbeta\ $\sim$ 0.23 given by 
Contini \etal\ \cite*{CONTINIetal97} instead of that derived by VC92. 
Our W-R region is indeed better identified in the 
spectroscopic observations of Contini \etal\ \cite*{CONTINIetal97}. 
For Tol~89, we adopt the 
reddening coefficient \cbeta\ $\sim$ 0.18 estimated by Terlevich \etal\ 
\cite*{TERLEVICHetal91} which is in good agreement with the measurements 
of Durret \etal\ \cite*{DURRETetal85}. 
See Schaerer \etal\ \cite*{SCHAERERetal97} for  NGC~5253. 

Dereddened fluxes are derived using the extinction law of Cardelli 
\etal\ \cite*{CARDELLIetal89} and including a Galactic foreground 
extinction $E(B-V)$ given in Table~\ref{GLOBAL}. 
The result is almost independent of the chosen extinction curve, since in 
the wavelength range $2600 - 9000$ \AA, there is little difference among the 
mean interstellar extinctions of the Galaxy, the Large Magellanic Cloud, and 
the Small Magellanic Cloud \cite{S79,BOUCHETetal85,F86}.
Dereddened line fluxes are given in Table~\ref{DATA}.


\begin{table*}[t]
  \caption[]{Relative flux and equivalent width of emission lines in the W-R 
regions of galaxies}
  {
\begin{tabular}{lrrrrrrr}
\hline
\hline
Line/Galaxy    & He~2-10 & NGC~3049 & \multicolumn{2}{c}{NGC~3125} & \multicolumn{2}{c}{NGC~5253} & {Tol~89} \\
\cline{4-5}
\cline{6-7}
        &         &          &  A & B                       & A & B                        &        \\
\hline
 & \multicolumn{7}{c}{\sc Nebular Emission Lines} \\
\hline
\heia 		 &          11$\pm$ 3&      11$\pm$ 3&	   42$\pm$  8&	   42$\pm$  3&	   49$\pm$  1&	    45$\pm$ 3&	   31$\pm$ 19\\
\feiiia		 &	    35$\pm$ 5&	    42$\pm$ 5&	   10$\pm$  3&	   23$\pm$  3&	   13$\pm$  4&	    19$\pm$ 2&	   27$\pm$  8\\
\ariva		 &\ldots	     &\ldots	     &	   13$\pm$  3&	   11$\pm$  1&	   14$\pm$  4&	    21$\pm$ 2&\ldots	     \\
\arivb		 &\ldots	     &\ldots	     &      4$\pm$  1&	    3$\pm$  1&	    8$\pm$  2&\ldots	     &\ldots	     \\
\heib		 &\ldots	     &\ldots	     &	    7$\pm$  2&	    8$\pm$  2&	    9$\pm$  1&	     9$\pm$ 1&	   11$\pm$  2\\
\oiiia		 &	   471$\pm$22&	   106$\pm$ 6&	 1970$\pm$ 67&	 1526$\pm$ 41&	 2070$\pm$ 36&	  1491$\pm$19&	 1229$\pm$ 33\\
\oiiib		 &	  1430$\pm$44&	   338$\pm$ 4&	 5852$\pm$197&	 4538$\pm$128&	 6155$\pm$104&	  4389$\pm$59&	 3636$\pm$111\\ 
\ni		 &	  12$\pm$   2&	    14$\pm$ 5&\ldots	     &      7$\pm$  2&	    9$\pm$  1&	     8$\pm$ 2&	    3$\pm$  1\\
\feiiib		 &\ldots	     &\ldots	     &\ldots	     &\ldots	     &	    5$\pm$  1&	     8$\pm$ 1&\ldots	     \\
\cliiia		 &\ldots	     &\ldots	     &	    6$\pm$  1&\ldots	     &	    4$\pm$  1&	     5$\pm$ 1&\ldots	     \\
\cliiib		 &\ldots	     &\ldots	     &\ldots	     &\ldots	     &	    3$\pm$  1&\ldots	     &\ldots	     \\
\nii		 &	    10$\pm$ 2&       8$\pm$ 1&\ldots	     &\ldots	     &	    5$\pm$  1&\ldots	     &\ldots	     \\
\heic		 &	   123$\pm$ 7&	   133$\pm$ 2&	  121$\pm$  3&	  120$\pm$  6&	  131$\pm$  2&	   130$\pm$ 1&	  118$\pm$  3\\
\oia		 &	    14$\pm$ 5&	    23$\pm$ 1&	   18$\pm$  3&	   19$\pm$  1&	   27$\pm$  5&	    28$\pm$ 2&	   13$\pm$  1\\
\siii		 &	    21$\pm$ 7&	     6$\pm$ 1&	   14$\pm$  1&	   12$\pm$  1&	   29$\pm$  7&	    20$\pm$ 1&	   20$\pm$  1\\
\oib		 &	     4$\pm$ 1&	     7$\pm$ 1&	    5$\pm$  1&	    7$\pm$  1&	    7$\pm$  1&	     8$\pm$ 1&	    7$\pm$  4\\[0.3cm]
$I$(\hbeta)	 &	   196$\pm$ 8&	    16$\pm$ 3&	   66$\pm$ 11&	   73$\pm$  8&	  212$\pm$  7&	    49$\pm$ 2&	   13$\pm$  1\\
$W$(\hbeta) [\AA]&	    23$\pm$ 3&	    34$\pm$ 3&	   93$\pm$  9&	   70$\pm$  7&	  269$\pm$ 23&	   112$\pm$ 10&	   68$\pm$ 17\\
\cbeta\		 &0.56&0.23&0.40&0.64&0.44&0.20&0.18\\
(O/H) [\zsun]	 & 0.42 & 1.20 & 0.17 & 0.20 & 0.20 & 0.20 & 0.25 \\
\hline
 & \multicolumn{7}{c}{\sc Broad Emission Lines from Wolf-Rayet Stars} \\
\hline
\niii +\Ciii	 &	   112$\pm$43&	   117$\pm$35&	   31$\pm$  4&	   63$\pm$ 18&	    7$\pm$  2&	    18$\pm$ 8&	   81$\pm$ 32\\
\heiia		 &	    97$\pm$35&	   128$\pm$38&	   76$\pm$  8&	   73$\pm$ 15&	    9$\pm$  3&	    45$\pm$15&	  116$\pm$ 43\\
\ciii		 &\ldots	     &	    35$\pm$10&\ldots	     &\ldots	     &\ldots	     &\ldots	     &\ldots	     \\
\civ		 &	    41$\pm$ 7&	    79$\pm$10&	   21$\pm$  5&	   51$\pm$  9&	    6$\pm$  3&	    30$\pm$ 5&	  132$\pm$ 30\\[0.3cm]
$W$(\niii +\Ciii) [\AA]&	  2.7$\pm$0.5&    5.2$\pm$0.5&	  3.2$\pm$0.5&	  2.8$\pm$1.0&	  2.1$\pm$1.3&	  1.0$\pm$0.5&	  3.4$\pm$0.5\\
$W$(\heiia) [\AA]&	  2.3$\pm$0.5&	  5.6$\pm$0.5&	  7.3$\pm$1.0&	  2.8$\pm$0.5&	  3.0$\pm$0.7&	  3.3$\pm$1.1&	  6.1$\pm$0.5\\
$W$(\ciii) [\AA]&\ldots	     &	  2.0$\pm$0.5&\ldots	     &\ldots	     &\ldots	     &\ldots	     &\ldots	     \\ 
$W$(\civ)	 [\AA]&	  1.5$\pm$0.5&	  4.7$\pm$0.6&	  3.1$\pm$0.4&	  6.2$\pm$1.0&	  2.5$\pm$0.6&	  6.0$\pm$1.1&	 11.9$\pm$1.5\\
\hline
\hline
\noalign{\smallskip}
\noalign{\small For each emission-line we reported the dereddened (\ilam) 
flux normalized to $I(H\beta) = 1000$. $I$(\hbeta) is the absolute dereddened 
flux of the \hbeta\ emission line in $10^{-14}$ \ergscm, \cbeta\ is the 
extinction coefficient (see Sect.~\ref{REDDENING} for references) and $W_{\lambda}$ 
is the equivalent width of the \hbeta\ and W-R emission lines in \AA. 
(O/H) is the oxygen abundance in solar units (see Sect.~\ref{OH} for details). 
}
\end{tabular}
  }
  \label{DATA}
\end{table*}


\subsection{Oxygen abundance}
\label{OH}

Oxygen is an important element for our subsequent analysis, as it is used to 
define the metallicity of each W-R region for the comparison with starburst 
models (see Sect.~\ref{MODELS}). Since the \oii\ and [O\,{\sc iii}] 
$\lambda$4363 lines are not included in our wavelength range, 
the oxygen abundance was determined from the empirical calibration between 
log(O/H) and the ratio $R_3$ \cite{EP84}, with the linear fit given by VC92. 
Edmunds \& Pagel \cite*{EP84} estimate the intrinsic uncertainty in the 
calibration to be about $\pm 0.2$ in log(O/H).
For the solar oxygen abundance, we adopted the 
local galactic value of log(O/H)$_{\odot} = -3.08$ \cite{M85}. 
The oxygen abundances derived in this way are summarized in Table~\ref{DATA}.
Except NGC 3049 which has an oxygen abundance slightly above solar, 
the values for the other W-R regions are below solar, typically 
$0.2 - 0.5$ \zsun.

Our derived oxygen abundances are generally in good agreement with 
earlier studies.
The oxygen abundance (O/H) $\sim$ 0.2 \zsun\ derived for NGC~5253 
is identical to the value obtained by Kobulnicky \etal\ \cite*{KOBULNICKYetal97}. 
We obtain (O/H) $\sim$ 1.2 \zsun\ for the W-R region of NGC~3049, 
in good agreement with previous observations by Contini \etal\ 
\cite*{CONTINIetal97}. A good agreement is also found for NGC~3125 between 
our measurements (0.17 \zsun\ for region $A$ and 0.20 \zsun\ for region $B$) 
and those of VC92. 
The same is true for Tol~89, when we compare our estimation of (O/H) $\sim$ 
0.25 \zsun\ to the derivation by Durret \etal\ \cite*{DURRETetal85}. 

For He~2-10, however, we obtain (O/H) $\sim$ 0.42 \zsun, three 
times larger than the value from VC92 for the 
same region. 
The calculations by VC92 are based on 
the detection of the temperature-sensitive [O\,{\sc iii}] $\lambda$4363 
emission line, which is however very faint in their spectrum of He~2-10. 
Their (O/H) value may therefore be underestimated, and we thus 
prefer the higher value derived from our spectra. 

\subsection{Broad emission lines from Wolf-Rayet stars}
\label{W-RSTARS}

Several broad emission lines (FWHM $\ga$ 25 \AA) due to W-R stars have been 
detected with high confidence level ($\ga$ 3 $\sigma$)  over the wavelength range 
covered by our spectra: the blend of 
\niii\ and \Ciii, 
\heiia, and \civ.
Equivalent widths and fluxes relative to \hbeta\ of these emission 
lines are listed in Table~\ref{DATA}. 

One of the main results of this paper is the unambiguous detection of  
broad (FWHM  $\sim 50 - 90$ \AA) \civ\ emission in {\em all W-R regions of 
the observed galaxies} 
(see Figs.~\ref{SPECALL} and \ref{SPEC5800}), 
which clearly indicates the presence of WC stars in these regions (cf.\ below).
Whereas all the objects where previously known to show the broad W-R bump around 
4700 \AA, our spectra are the first to show the red W-R bump around 5800 \AA.
As already pointed out by Schaerer \etal\ \cite*{SCHAERERetal97}
a sufficiently high signal-to-noise ratio ($\ga$ 40) is required 
for the detection of such a  \civ\ line in integrated spectra of W-R galaxies.
Typically the 5808 flux is found to be factor of two less than 4686;
furthermore the larger width of \civ\ and the lower intensity of 
the continuum render its detection more
difficult.
This explains why this line has remained undetected in previous ground-based
 spectroscopical studies, except for He~2-10 
\cite{HS84,VC92}. 

What do the broad emission features tell us about the W-R populations
in these regions ?
WN stars cannot be responsible for the \civ\ emission, since they show
\heiia/\civ\ $\sim$ 16. (SV98), much larger than observed in our spectra.
In addition the strong \heiia\ emission, the relative strengths
of the W-R features in the blue bump, 
and the FWHM(\heiia) ($\sim 30 - 40$ \AA) are fingerprints of the presence of 
WN stars, as commonly accepted \cite{C91}.
In all our seven emission line regions we thus find signatures of
both WN and WC stars. For six regions this represents the first detection 
of WC stars.

The dominant W-R subtypes can be constrained as follows.
In all our spectra the presence of \niii\ and/or \Ciii\ is established,
while \nv\ is very weak or absent. If the former are due only to WN stars 
this indicates the predominance of late-type WN stars (WNL).
The relative strength of (\niii+\Ciii)/\heiia\ ranges from 0.9 to 2.5 with an
average of 1.5, which is much larger than the average of $0.24 \pm 0.15$ 
obtained for LMC WNL stars (SV98). The excess flux can easily be attributed to 
the emission from the WC stars.
As for the WN stars it is not possible to apply the usual criteria to 
classify the WC subtypes. However, the measured FWHM(5808) are compatible 
with types WC7-WC4 but also WO1, WO3, and WO4 \cite{SMITHetal90,CROWTHERetal98}. 
A significant population of WO1 stars can, however, be excluded from 
the absence of \ov. 
Only in NGC~3049 do we also detect \ciii\ (cf.\ below) with a relative
intensity compatible with WC7 stars.  In the other cases the non-detection
of \ciii\ and FWHM(5808) favors early-type WC stars. 
From the comparisons of the broad emission features, we thus conclude that
all regions (except the high metallicity galaxy NGC~3049 discussed below) 
contain a mixed population of WNL, and early WC and/or WO3-4 stars.
The presence of intermediate-type WN/WC stars cannot be excluded.

In the following we shall now discuss the W-R signatures individually 
for each object.


\begin{table*}
\caption[]{Massive star populations and mechanical energy input rates in W-R 
regions}
\begin{flushleft}
{\footnotesize
\begin{tabular}{llcrrrcrccrr}
\hline
\hline
Galaxy &  & Age & $Q_0^{\rm obs}$ & $N_{\rm WN}$ & $N_{\rm WC}$ & $\eta_0$ & $N_{\rm O}$ 
       & log $M_{\star}$ & log $\dot{E}$ & $r_{\rm b}$ & $v_{\rm b}$ \\
         &  & [Myr]     &  [$10^{49}$ s$^{-1}$]    &         & & & & [\msun] & [\ergs] & [pc] & [\kms] \\
(1) & & (2) & (3) & (4) & (5) & (6) & (7) & (8) & (9) & (10) & (11) \\
\hline
He 2-10  &  & 5.5 -- 6.0 & 3802 & 1100$\pm$520 & $>$ 250 & 0.5 -- 1.0  & 2450 - 4900 & 6.8 & 40.81 &1000 &110 \\
NGC 3049 &  & 5.5        & 1349 & 510$\pm$240 & $>$ 170  & --$^a$      & 0$^a$       & 6.4 & 40.40 & 900 &100 \\
NGC 3125 & A& 4.5 -- 5.0 & 2188 & 500$\pm$230 & $>$  70  & 0.25 -- 0.5 & 3240 - 6470 & 6.1 & 40.48 & 700 & 90 \\
         & B& 4.5 -- 5.0 & 2455 & 530$\pm$250 & $>$ 200  & 0.25 -- 0.5 & 3450 - 6900 & 6.1 & 40.53 & 700 & 90 \\
NGC 5253 & A&    3.0    &  896  &  26$\pm$13  & $>$   9  & 0.8  -- 0.9 & 960  - 1080  & 6.6 & 39.00 & 230 & 50 \\
         & B& 5.0       &  207  &  27$\pm$13  & $>$  10  & 0.25        & 680          & 6.6 & 39.30 & 440 & 50 \\
%
Tol 89   &  & 4.5 -- 5.0 &  708 & 240$\pm$110 & $>$ 150  & 0.25 -- 0.5 & 640  - 1270 & 5.7 & 39.84 & 550 & 70 \\
\hline
\hline
\noalign{\smallskip}
\noalign{\small $^a$ For instantaneous burst models with solar metallicity and higher, $\eta$ is not defined.
The derived number of O stars (col.\ 7) is strongly dependent on the adopted temperature limit between
O and B stars. See text for more details.}
\end{tabular}
}
\end{flushleft}
\label{STARPOP}
\end{table*}

\subsubsection*{He~2-10}
He~2-10 was the first galaxy discovered to have evidence 
for W-R stars \cite{ALLENetal76}, and is in many ways considered as 
the prototypical W-R galaxy \cite{C91}. A broad feature of \civ\ was 
also suspected by Hutsemekers \& Surdej (1984) and detected by VC92
in the optical spectrum of He~2-10. 
The presence of this line is unambiguous in our spectrum of 
He~2-10 (see Fig.~\ref{SPEC5800}). The blue W-R bump 
(see Fig.~\ref{SPEC4700}) is composed by the broad \heiia\ 
line and \niii +\Ciii\ with nearly the same strength. 

\subsubsection*{NGC~3049}
In NGC~3049, the W-R region first noted by Kunth \& Schild \cite*{KS86} 
is located in the bar, $\sim$ 2.5\arcsec\ North-East of the galaxy nucleus 
(see Fig.~\ref{N3049_SPAT}). The blue W-R bump is defined by two 
broad emission lines identified as \heiia\ and \niii + \Ciii\ (see 
Fig.~\ref{SPEC4700}), with approximatively the same strength as
already reported in previous observations (Kunth \& Schild 1986, VC92).
In addition 
to the \civ\ line, a broad (FWHM $\sim$ 45 \AA) emission line of \ciii\ is 
also detected at $\sim 2 \sigma$ level in the spectrum of NGC~3049 (see 
Fig.~\ref{SPEC5800}). 
After Phillips \& Conti~\cite*{PC92} this is the second detection of this 
W-R emission line in an extra-galactic \hii\ region.
The C~{\sc iii} emission is attributed to late WC stars, whose presence
is indeed expected in high metallicity regions \cite{M91}.

\subsubsection*{NGC~3125}
Penston \etal\ \cite*{PENSTONetal77} detected the \heiia\ emission line 
in the North-West knot (= our region $A$) of NGC~3125, but do not mention
W-R stars. The W-R nature of NGC~3125 was first noted by 
Kunth \& Sargent~\cite*{KS81} who first envisioned their importance to elucidate the
bursting nature of star-formation processes in \hii\ galaxies. They computed that this
 starburst region contains 
several hundred W-R stars. Long-slit observations of 
VC92 have first revealed
W-R stars in the second condensation 10\arcsec\ South-East of the nucleus 
(= our region $B$). In region $A$ of NGC~3125, the W-R bump around 4700 \AA\ 
is dominated by the \heiia\ line whereas the  \niii + \Ciii\ blend
and the \heiia\ line are of similar strength in region $B$ (see 
Fig.~\ref{SPEC4700}). 
We signal a marginal ($\sim 1 \sigma$) detection of broad He\,{\sc ii} 
$\lambda$5412 emission, the second strongest He\,{\sc ii} line in WN stars,
in knot A.
If confirmed this represents, to our knowledge, the first detection of this 
line in the integrated spectrum of an extra-galactic object.
The spectrum of region $A$ also exhibits the typical high-excitation nebular 
emission lines of \ariva\ and \arivb\ (see Fig.~\ref{SPEC4700}). 
Their presence might imply a nebular contribution to the \heiia\ line.

\subsubsection*{NGC~5253}
The spectra are discussed in Schaerer \etal\ \cite*{SCHAERERetal97}.
In addition we signal that He\,{\sc ii} $\lambda$5412 emission may 
also be marginally detected in region B.

\subsubsection*{Tol~89}
Durret \etal\ \cite*{DURRETetal85} reported the first detection of the W-R 
bump around 4700 \AA\ in the spectrum of Tol~89, but their low 
signal-to-noise ratio and moderate spectral resolution did not allow a 
clear identification of the lines. 
They do not detect any optical spectral signature from WC stars, 
but their presence was suggested from UV P-Cygni lines.
In our spectrum of Tol~89 (see Fig.~\ref{SPEC4700}), the blue W-R bump 
shows \heiia\ and at a lower level the \niii + \Ciii\ blend. 
The \civ\ line is the strongest ($W \sim$ 13 \AA) and the broadest 
(FWHM $\sim$ 90 \AA) detected among the spectra of observed W-R regions 
(see Fig.~\ref{SPEC5800}).

\subsection{Massive star populations}
\label{MASSPOP}

In this section we derive the approximate number of massive stars 
(O and W-R) in the W-R regions of galaxies. 
The mechanical energy output from the stellar population and its impact 
on the surrounding ISM is studied in Sect.~\ref{ENERGY}. 
A direct comparison of the observed W-R emission line features with
evolutionary synthesis models is presented in Sect.~\ref{MODELS}.

The approximate number of WN, WC, and O stars are listed in 
Table~\ref{STARPOP} (columns 4, 5 and 7). 
The number of W-R stars is calculated from the luminosity of the W-R 
emission lines. We assume that the dominant contributors to the broad 
\heiia\ and \civ\ lines are respectively WNL and WC4 stars 
(see Sect.~\ref{W-RSTARS}). 
The average observed luminosity of WNL stars in the \heiia\ line is
$L_{\rm 4686} = 1.6\pm 1.5 \times 10^{36}$ \ergs; that of WC4 stars in the 
\civ\ line is $L_{\rm 5808} = 3.0\pm 1.1 \times 10^{36}$ \ergs (SV98). 
The quoted uncertainty on the WN number reflects the standard deviation
of $L_{\rm 4686}$. 
Given the complex structure of the blue W-R bump, the uncertainty
on \heiia\ is probably larger than the formal errors, and a nebular He\,{\sc ii}
contribution may also be present in some of our objects (NGC~3125-A,
NGC~5253-A, and possibly also Tol 89; see Fig.\ \ref{SPEC4700}). 
In this case $N_{\rm WN}$ has to be corrected downwards.
Since $L_{\rm 5808}$ is lower for later WC types (SV98) and the 
dominant WC type is not well constrained (see Sect.~\ref{W-RSTARS}) a lower 
limit on $N_{\rm WC}$ is provided.
Taking into account the variation in $L_{\rm 5808}$ between WC4-6 stars
we conclude that the number of WC stars may be larger by a factor of 
$\sim 3 - 5$. 

Under the condition of case B recombination and assuming that all the 
ionizing photons emitted by the stars are absorbed by the gas, the total 
number of Lyman photons, $Q_0^{\rm obs}$ (col.\ 3), can be derived from the observed 
luminosity of the \hbeta\ emission line. 
Note that this also provides a convenient estimate of the number of 
equivalent O7V stars, for which $Q_0^{\rm O7V} =  1.0 \times 10^{49}$ s$^{-1}$ \cite{L90}.
%
To estimate the number of O stars, $N_{\rm O}$, we must take into account 
the ionizing photon contribution from W-R stars, the age of the stellar population 
and the IMF (see SV98). $N_{\rm O}$ is given by:
\begin{equation}
N_{\rm O} = \frac{Q_0^{\rm obs} - N_{\rm W-R} Q_0^{\rm W-R}}{\eta_0(t) 
	    \times Q_0^{\rm O7V}},
\end{equation}
where $N_{\rm W-R} = N_{\rm WN} + N_{\rm WC}$, $Q_0^{\rm W-R}$ is the 
average Lyman continuum photon flux per W-R star, and $\eta_0(t)$ 
is the IMF averaged ionizing Lyman continuum luminosity of a stellar population 
normalized to the output of a O7V star (see SV98).
In most regions we find 
$N_{\rm W-R}/N_{\rm O7V} = (N_{\rm WN}+N_{\rm WC})/N_{\rm O7V} \sim 0.2 - 0.5$.
Assuming a priori that the massive main sequence stars provide the bulk
of the ionizing Lyman continuum photons implies that the average Lyman flux
per W-R star, $Q_0^{\rm W-R}$, is not significantly larger than $Q_0^{\rm O7V}$.
For simplicity we adopt $Q_0^{\rm W-R} = Q_0^{\rm O7V} = 1.0 \times 10^{49}$ 
s$^{-1}$ (similar to VC92 but smaller than the age dependent value given by SV98).
The value of $\eta_0$ given in column 6 of Table~\ref{STARPOP} was taken from the 
instantaneous burst models of SV98 for a Salpeter IMF at the age given in column 2
(derived from $W(\hbeta)$). The resulting number of O stars (defined by
$T_{\rm eff} >$ 33\,000 K) is given in column 7. The indicated range results only
from the range of $\eta_0$. The number of O stars found for the
W-R regions are between $\sim$ 500 and 7000. 
NCG 3049 is somewhat exceptional in this respect. Indeed at its age (determined
from $W(\hbeta)$) the instantaneous burst models at solar or higher metallicity
predict that no stars with $T_{\rm eff} >$ 33\,000 K are left.
If this temperature limit is adopted to separate O stars from later types (cf.\ 
SV98) we formally derive $N_{\rm O}=0.$ (see Table \ref{STARPOP}).
Lowering this limit by 10 \% already changes the situation qualitatively:
in this case one obtains $\eta \sim 0.16$, and $N_{\rm O} \sim$ 4000 as test
calculations show.
In any case, according to these models, the bulk of the ionisation is provided 
by late O and/or B type stars and W-R stars.


From Table~\ref{STARPOP}, we obtain W-R/O number ratios of $\sim 0.03 - 0.6$
(and even larger for NGC~3049), systematically higher than the predictions 
for constant star formation 
at the appropriate metallicity \cite{MM94}, but within the range of instantaneous
burst models with different IMF slopes (SV98, Table 4).
A more detailed attempt to constrain the IMF is made in Sect.\ \ref{IMF}.
A trend of increasing W-R/O ratios towards higher metallicity is found
as expected on the average (e.g.\ Meynet 1995).

For the seven W-R regions, we find a flux ratio for \heiia/\civ\ of $1.8\pm 0.8$,
which shows a fairly small dispersion.
The corresponding number ratio of WC/WN stars is typically $0.2 - 0.4$; the extreme
values are 0.14 (NGC~3125-A) and 0.63 (Tol 89). No systematic variation with 
metallicity is found. 
For the low-metallicity objects ($Z \sim$ 0.2 \zsun; NGC 3125, NGC~5253, and 
Tol~89) the derived WC/WN ratio is larger than what is found in the Local Group 
(except IC~10) at similar metallicity \cite{MJ98}. 
At higher metallicities He~2-10 and NGC~3049 show, however, WC/WN ratios below the 
trend observed by Massey \& Johnson~\cite*{MJ98}.

\begin{figure}[t]
\psfig{figure=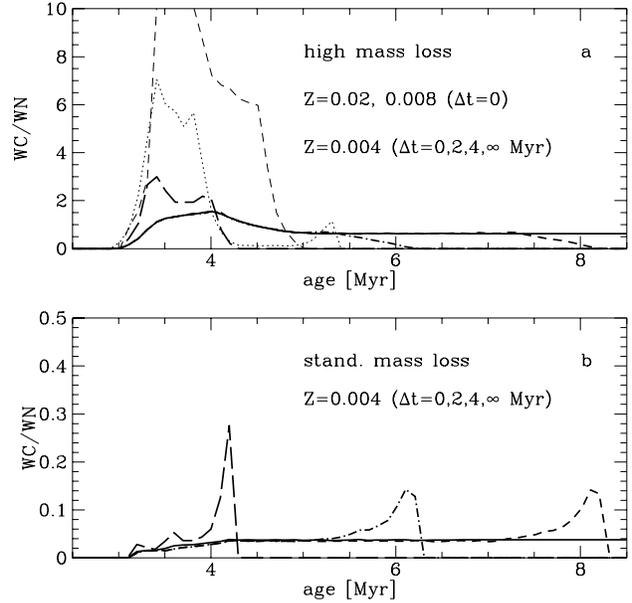,width=8.8cm}
\caption{Predicted WC/WN number ratio as a function of age for 
stellar evolutionary tracks assuming high mass loss rates ({\bf a}) and 
standard mass loss rates ({\bf b}) as described in Meynet \etal\ (1994).
Instantaneous burst models for solar metallicity (Z=0.020, thin
dashed line) and Z=0.008 (thin dotted) are shown in {\bf a}.
Thick lines show models at Z=0.004 for different burst durations:
instantaneous burst (long-dashed), $\Delta t=2$ Myr (dashed-dotted),
$\Delta t=4$ Myr (dashed), and constant star formation (solid).
All models assume a Salpeter IMF.
\label{WCWN}}
\end{figure}

The finding of a fairly constant WC/WN ratio may seem surprising at first sight.
Contrary to regions of constant star formation, likely representative of the Local
Group samples, regions of short star formation (more appropriate for our observed
starburst galaxies, see Sect.~\ref{IMF}) could a priori show quite a large range 
in WC/WN, depending on the age of the starburst. 
This is illustrated in Fig.~\ref{WCWN} where the predicted WC/WN ratio from the 
SV98 models are shown for different metallicities, evolutionary tracks, and burst 
durations (cf.\ also Meynet, 1995\footnote{As pointed out in SV98 due to a 
different interpolation technique the 
models of Meynet~\cite*{M95} predict a more important WC population at later ages than 
SV98 even if the same tracks are used. The ``real'' number of WC stars is likely 
intermediate between the two models. For a given set of tracks the WC/WN plotted in 
Fig.~\ref{WCWN} may thus be somewhat underestimated.}).
However, already burst durations of $\ga 2$ Myr suffice to smooth out the rapid
variations of WC/WN. This may well explain the small range of WC/WN found
for the low-metallicity objects. 
The observed WC/WN value in these objects is also intermediate between the predictions 
from high and standard mass loss models (cf.\ Fig.~\ref{WCWN}a and \ref{WCWN}b). 
More surprising is the low WC/WN ratio of the remaining higher metallicity objects 
(He~2-10 and NGC~3049) compared to the WC/WN value of Massey \& Johnson \cite*{MJ98} 
at a similar metallicity. Indeed, the probability of finding WC/WN ratios below 
the equilibrium value attained in regions of constant star formation should be quite 
small.
However, if the number of WC stars is systematically underestimated by a factor of 3 
(due to variations of the average WC subtype with metallicity; cf.\ above) 
our observations may well all be larger than the observed WC/WN trend with
$Z$.
We conclude that the WC/WN ratios of the low-metallicity galaxies NGC~3125, NGC~5253 
and Tol~89 can be understood quantitatively with burst models of reasonably short 
but non-zero duration. 
Additional observations of WC and WN populations, especially for regions of higher 
metallicities, would be very helpful.

\subsection{Energy released by massive stars}
\label{ENERGY}
The winds produced by these populations of massive stars and by supernovae 
explosions should strongly perturb the gas dynamics of galaxies. 
We effectively observe some perturbations centered on the W-R regions in 
the velocity curves of the ionized gas (see Figs.~\ref{HE0210_SPAT} to 
\ref{TOL89_SPAT}). The observed velocity differences range from $\sim$ 40 
\kms\, in NGC~3125 and NGC~5253 to $\sim$ 100 \kms\ in He~2-10 and NGC~3049. 
These values are similar to the typical expanding velocities 
($\sim$ 50 \kms) of superbubbles and filaments observed in dwarf 
galaxies \cite{MARLOWEetal95,LK97}. According to these authors, the mechanical 
energy output from the supernovae explosions and strong stellar winds in 
the starburst regions appears adequate to power expansion motions of this 
speed. Note also that gas flows resulting from the energizing of the interstellar
medium are currently observed in star-forming galaxies \cite{K98}. 

To verify this claim, one can compare the computed expansion velocity 
of a bubble of gas centered on the W-R regions to the observed velocity 
differences in galaxies.
We used a very simple model of wind-blown bubble expanding adiabatically 
(i.e. energy conservative), which has been described by Castor \etal\ 
\cite*{CASTORetal75} who give the following expressions for the evolution of 
the bubble radius (in kpc) and expansion speed (in \kms):
\begin{equation}
r_{\rm b} = \left(\frac{\dot{E}_{41}}{n_0}\right)^{1/5} \times t_7^{3/5}
\end{equation}     
\begin{equation}
v_{\rm b} = 62 \times \left(\frac{\dot{E}_{41}}{n_0}\right)^{1/5} \times t_7^{-2/5}
\end{equation}
where $\dot{E}_{41} \equiv (dE/dt)_{41}$ is the kinetic injection rate in 
units of 10$^{41}$ \ergs, $n_0$ is the number density (in \cmc) in the 
ambient interstellar medium, and $t_7$ is the time since the expansion of 
the bubble began in units of 10$^7$ yr. The rate of kinetic energy produced 
by the population of massive stars $\dot{E}_{41}$ can be estimated 
from starburst model predictions \cite{L&H95} assuming an instantaneous 
burst and a Salpeter IMF. The predicted normalized rate of mechanical energy 
at a given age is thus multiplied by the total mass of ionizing stars formed in 
the starburst region. The total burst mass (column 8 of Table~\ref{STARPOP}) 
is derived from $Q_0^{\rm obs}$ using the instantaneous burst models of SV98 and 
assuming a Salpeter IMF down to 0.8 \msun. 
We adopt $n_0 \sim 0.3$ \cmc\ \cite{MARLOWEetal95}, and 
for $t_7$, we use the estimated age of the starburst listed in 
Table~\ref{STARPOP} (column 2). The computed radius $r_{\rm b}$ and expansion 
velocity $v_{\rm b}$ of the bubble are given in the last two columns of 
Table~\ref{STARPOP}. 

Comparing these values to the velocity differences 
observed in the W-R regions of galaxies (from $\sim$ 40 to 100 \kms), 
we find that the observed and predicted bubble expansion speeds are in 
satisfactory agreement, especially when considering the extreme simplicity 
of the model and the observational uncertainties. 
The comparison of radius is more doubtful since, with 
long-slit spectra, we have some informations in only one spatial direction. 
The rough assessments of ``bubble'' radius from velocity curves are 
systematically lower by a factor two (NGC 3049) to six (He~2-10) than the 
values predicted by the model. The same trend has been observed using 
\halpha\ images of dwarf galaxies \cite{MARLOWEetal95}. These discrepancies 
are certainly due to the difficulty to measure a ``radius'' from the velocity 
curves and to the simplicity of the model which assumes a pure spherical 
bubble expanding adiabatically. When looking deeply in \halpha\ images of 
nearby galaxies, like NGC~5253 \cite{CALZETTIetal97}, one can see that the 
structure of expanding gas is much more chaotic, ranging from large scale 
($\sim$ 1 kpc) filaments to roughly circular shell nebulae of smaller 
size ($<$ 100 pc). It is thus very difficult and even meaningless to compare 
some predicted and observed bubble ``radius''. 

Nevertheless, the size ($\sim$ 50 to 300 pc) of expanding gas structures observed in 
W-R regions is significantly smaller than those derived from \halpha\ 
images ($\sim$ 700 
to 1500 pc) of dwarf galaxies \cite{MARLOWEetal95}. This might reflect the 
different time scales of the gas dynamics, which depend on the age of the 
starburst responsible for the release of kinematical energy through 
stellar winds and supernovae explosions. In the W-R regions, the starburst 
is one order of magnitude younger (a few $10^6$ yr) than the one (a few 
$10^7$ yr) derived from gas kinematics in older starburst regions 
\cite{MARLOWEetal95}. This could explain the smaller sizes of expanding 
``bubbles'' in W-R regions. 
An additional effect may be the smaller energy injection expected from young
regions where the SN rate is still small \cite{L&H95}.
In any case it is evident that the large population of massive stars 
derived in W-R regions strongly perturb the surrounding ISM through the injection 
of their mechanical energy.
 
\section{Comparison with starburst models}
\label{MODELS}

All the observed W-R features and the most important nebular H and He 
lines have been modeled recently with synthesis models by SV98.
The aim of this Section is to constrain the main burst parameters
(age, IMF, star formation history) as far as possible from the observed 
W-R signatures. Finally the success or failure to reproduce the observed features
also provides a test of the underlying evolutionary tracks.

\subsection{Model parameters and comparison procedure}
\label{s_procedure}

In the following we consider two basic free model parameters: 
the IMF slope and the duration of the star-forming event.
The third model parameter, the metallicity $Z$, is adopted from
the observations (see values given below).
A power law with a slope $\alpha$ is adopted for the IMF 
(in our notation $\alpha = 2.35$ for a Salpeter IMF). 
The upper mass cut-off is generally set to 
$M_{\rm up}=$ 120 \msun; the results discussed in this work are not
affected by the choice of the lower mass cut-off.
We consider burst models with different durations $\Delta t$ starting 
at time $t=0$, with the limiting case of an instantaneous burst
($\Delta t=0$).

For the subsequent detailed comparison it is, however, useful to remind
some assumptions or uncertainties of the models, which are not considered
as proper free parameters.
\begin{itemize}
\item[{\em (a)}] Most important is the set of stellar evolution models adopted. We use 
SV98 models based on the latest Geneva stellar evolution tracks (high mass 
loss models of Meynet \etal, 1994), which have been extensively compared
to observations (see Maeder \& Meynet 1994) and which in particular reproduce
well massive star populations in the Local Group.
Only single star models are considered (but see Sect.\ 5.3 for
binary stars).

\item[{\em (b)}] The predicted fraction of WC stars (WC/WR, and WC/WN)
is affected by the choice of interpolation techniques as briefly discussed
in Sect.\ 4.5 and SV98. During the WR phase at ages $t \ga$ 4 Myr the relative 
WC populations predicted by the Meynet (1995) and SV98 models differ; compared 
to the models used here (SV98), more WC stars are likely to exist at $t \ga$ 4 Myr.

\item[{\em (c)}] As mentioned in Sect.\ \ref{MASSPOP} the predicted strength of
the W-R lines is affected by the uncertainty of the intrinsic W-R line luminosity,
which is the largest for \heiia.

\item[{\em (d)}] To relate models with observations $W(\hbeta)$ is conveniently used
as an age indicator. This assumes that the models predict the correct
number of ionizing photons (i.e.\ correct atmospheres, tracks), Case B
(ionization bounded nebula) is valid and the fraction of photons absorbed by
the gas is correct (SV98 adopt $f_\gamma=1$), and the continuum light
is correctly predicted. All except the last point determine obviously
the correctness of the nebular line intensity, whose strength is also used
for comparisons with the WR lines (see below).
It is fairly difficult to verify how well the predicted $W(\hbeta)$--age
relation (cf.\ Copetti \etal 1986) holds. Empirical tests using recent models 
and observations of \hii\ regions and their stellar content in the Galaxy and 
the Magellanic Clouds could be useful. The fact that very few \hii\ regions
with observed $W(\hbeta)$ as large as predicted by synthesis models for ages
$\sim$ 0--2 Myr may indicate a difficulty in the $W(\hbeta)$--age relation
(e.g.\ Mas-Hesse \& Kunth 1998).

\end{itemize}

The comparison with the observations can be performed in two ways.
One may compare relative line intensities of the W-R features with
respect to the nebular \hbeta\ emission (in short W-R/\hbeta), or use 
equivalent widths of the W-R lines. 
Observationaly the determination of these quantities can mainly be ``perturbed'' 
by three effects:
\begin{itemize}
\item[{\em (i)}] 
We do not count all the ionizing photons produced by massive stars.
  This may happen if the nebular emission from the \hii\ region of interest
  is not entirely included in the slit (e.g. if the gaseous component is
  more extended than the ionizing star clusters), and/or if
  ionizing photons can escape from the \hii\ region (i.e.\ not
  ionization bounded nebula).
\item[{\em (ii)}] Stars and gas suffer from a different extinction.
\item[{\em (iii)}] An underlying older population contributes additional 
	continuum light.
\end{itemize}
The relative intensities W-R/\hbeta\ are effected only by  {\em (i)} and 
{\em (ii)}.
Effect {\em (i)} increase the relative WR/\hbeta\ intensities.
The same holds for  {\em (ii)} if the stellar light is less extincted
than the gas (cf.\ e.g.\ Calzetti 1997, Mas-Hesse \& Kunth 1998).
The equivalent widths are only affected by {\em (iii)}, which decreases 
the observed value.
For the present paper the {\em quantitative} importance of these effects 
cannot be asserted in general without additional observations. 
At least our spatial information (see Sect.~\ref{W-RDISTRI}) can provide hints 
on {\em (i)}.
From the work of Mas-Hesse \& Kunth (1998)
we see that, if present, the differences 
in extinction between the gas and stars are typically 
$\Delta E(B-V) \sim$ 0.05 -- 0.1.
This implies that the stellar flux may be overestimated by a factor of
$\sim$ 1.2 -- 1.4, i.e.\ $I({\rm W-R})/I(\hbeta)$, $N_{\rm WN}$ and
hence $N_{\rm WN}/$O have to be reduced by the same amount.
Although e.g.\ the WC/WN ratio is in principle also modified, the effect
is small ($\sim$ 4 -- 8 \%).
For NGC~3125, NGC~5253, and NGC~3049 an underlying population contributes to
$\sim$ 60 \% of the light at 4630 \AA\ (Mas-Hesse 1998, private communication),
which would correspond to a downward correction by a factor of $\sim$ 1.7
for e.g.\ $W(4686)$.
In our observations the contribution from an underlying population is, however,
most certainly smaller since they are taken with smaller apertures
including the bulk of the young emission line regions (cf.\ also 
Schaerer \etal 1997) but only a smaller fraction of the continuum light.

Given these potential difficulties both comparisons of relative line intensities
W-R/\hbeta\ and W-R equivalent widths will be performed for all objects. 
Significant differences between the two methods likely indicate some difficulty
with {\em (d)} and/or effects {\em (i)--(iii)}.
In this case the simplest meaningful comparison is between the observed 
WR equivalent widths and the maximum value predicted by the models 
(irrespectively of the age); it only depends on {\em (iii)}.

It is well known that the W-R populations strongly depend on
metallicity. We will therefore compare the observed regions 
to the models at the closest metallicity available. For the
remainder of this Section we will associate the objects 
(O/H given in Sect.~\ref{OH}) with the following metallicity $Z$:
NGC 3049: $Z \ga$ 0.02 (star symbol in Figs.\ \ref{fig_ihb} to
\ref{fig_ew_004}), He 2-10: $Z=$ 0.008 (open triangle), 
and $Z=$ 0.004 for the remaining objects (NGC 3125, NGC 5253, 
Tol 89; all filled symbols).

\subsection{Results}

\begin{figure}[t]      
\centerline{\psfig{figure=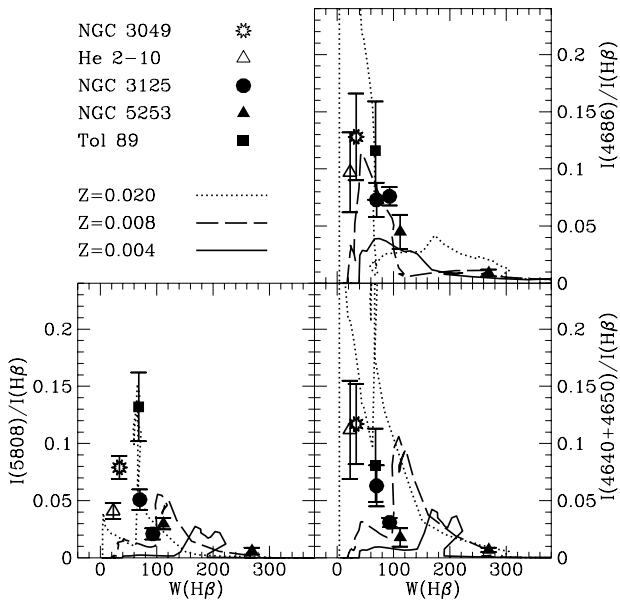,width=8.8cm}}
\caption{Observed and predicted W-R line intensities relative to
\hbeta\ as a function of the \hbeta\ equivalent width. 
The symbols used are show on the Figure. Filled symbols
designate objects with a metallicity of $Z \sim$ 0.004; open symbols
have a larger $Z$ (see text). The fairly small uncertainties on $W(\hbeta)$ 
are not shown in this Fig.\ (see Fig.\ \protect\ref{fig_ew_004}).
Model predictions from SV98
(instantaneous burst, Salpeter IMF) are shown for $Z=$ 0.020 (dotted),
0.008 (dashed) and 0.004 (solid).
Upper right: relative line intensity for \heiia;
lower left: \civ; lower right: \niii +\Ciii\ blend.
Discussion in text}
\label{fig_ihb}
\end{figure}

\begin{figure}[t]        
\centerline{\psfig{figure=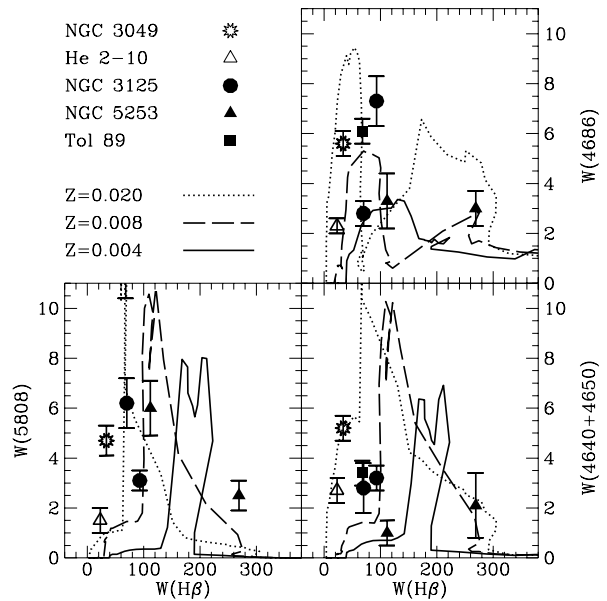,width=8.8cm}}
\caption{Same as Fig.\ \protect\ref{fig_ihb} for the W-R equivalent
widths. Note that the observed value of $W($\civ$)$ for Tol 89 exceeds the
scale shown here (see Table \protect\ref{DATA}).
Discussion given in text}
\label{fig_ew}
\end{figure}

In Fig.\ \ref{fig_ihb} we compare the observed relative W-R line 
intensities of all W-R regions with the model predictions of SV98
at $Z=$ 0.02 (solar), 0.008, and 0.004 for an instantaneous burst
with a Salpeter IMF (hereafter ``standard'' model; variations are
considered in Sect.\ \ref{IMF}). The same is shown for the W-R 
equivalent widths in Fig.\ \ref{fig_ew}. In both plots, $W(\hbeta)$ 
is used as a time indicator (decreasing $W$ with time) in the 
synthesis models.
The following general comments can be made to Figs.\ \ref{fig_ihb} 
and \ref{fig_ew}.

From object to object the observed line ratios vary by a much 
larger factor than the W-R equivalent widths. 
This can be explained by the more rapid decrease of the Lyman 
continuum luminosity (i.e.\ the \hbeta\ intensity) compared to
the smaller variations in the optical continuum light in a short
burst.
With the exception of \civ\ in Tol 89 (cf.\ below), larger line 
ratios are found in objects with higher $Z$. This is to first order 
understood by the more rapid decline of the ionizing luminosity
with increasing metallicity, whereas the increase of the W-R
population is less important (SV98).
These findings favor the picture of short burst periods (compared
to the lifetime of massive stars) and support the predicted
dependence of the ionizing population with metallicity found
independently from other studies \cite{GVetal95,SL96}. 

W-R signatures are observed at \hbeta\ equivalent widths where the 
models indeed predict W-R stars. Although this is not a strong 
constraint it indicates that the $W(\hbeta)$--age chronometer 
is reasonably synchronised to first order. 
Some difference seems, however, to be present for the appearance 
of WC signatures (see also Fig.\ \ref{fig_ew_004}). Given the 
uncertainty {\em (b)} in the models, this difference is not 
significant (see below).

Comparing the individual objects (classified by metallicity) with the 
model predictions, we find the following:

\subsubsection*{NGC 3049 ($Z \ga 0.02$)}
The predictions of the standard model (instantaneous burst, 
Sal\-peter IMF) for the high metallicity region
in NGC 3049 show an excellent agreement with all observed W-R
line strengths. This is also the case for the \ciii\ feature
(not shown here) which is only detected in NGC 3049.

\subsubsection*{He 2-10 ($Z = 0.008$)}
For He 2-10 the comparison between the equivalent widths and the 
line ratios shows a somewhat different picture. 
While the observed equivalent widths are considerably smaller than 
the maximum value predicted by the models (for the corresponding $Z = 0.008$), 
the relative line intensities are close to the predicted maximum.
This difference can be attributed to the mismatch between gaseous and 
stellar emission (cf.\  {\em (i)}) seen in Fig.\ \ref{HE0210_SPAT}, 
although other explanations (e.g.\ leakage of photons) cannot be ruled out.
Assuming that effect {\em (iii)} is small (cf.\ below) we conclude
from the comparison of the W-R equivalent widths (Fig.~\ref{fig_ew}) 
that the W-R and O star populations are compatible with an 
instantaneous burst and a Salpeter IMF.
The same conclusion was already
reached by Schaerer~\cite*{S96} based on the observations of VC92.

\subsubsection*{NGC 3125, NGC 5253 and Tol 89 ($Z = 0.004$)}
Only some of the observed W-R line intensities and equivalent widths
lie well on the predicted model curves plotted in Figs.\ \ref{fig_ihb}
and \ref{fig_ew}.
NGC~5253 represents the best case \cite{SCHAERERetal97}.
For the considered objects we dispose in total of 15 measurements 
of broad emission lines (all lines and all regions).
The relative line intensity of 7 data points is above the maximum
value predicted by the standard model. On the other hand the majority 
(12 of 15) of the equivalent widths are well in the range of the model 
predictions. The three remaining cases are $W($\heiia$)$ of
NGC~3125-A and Tol~89, which could well be overestimated due to nebular
contamination and contributions from other lines in the blended region,
and an exceptionally large \civ\ in Tol 89 (cf.\ below).

The uncertainties affecting the observed quantities have been
discussed above (Sect.\ \ref{s_procedure}).
Only for Tol~89 we have direct evidence that effect {\em (i)} 
(displacement gas--stars) is likely of importance (Sect.\ \ref{SPATANAL}); 
this does, however, not exclude that the line intensities in some 
other objects are also affected.
Even after reducing $I({\rm WR})/\hbeta$ by a factor $\sim$ 1.4
due to effect {\em (ii)}, discrepancies in the line intensities
remain.
The observed W-R equivalent widths are mostly within the predicted
range, although the observed values could be underestimated due to 
effect {\em (iii)}. We feel, however, that our measurements are only 
weakly affected by the old stellar populations (see Sect.\ \ref{s_procedure}).
Other possibilities to explain the observations of some regions with 
large W-R/\hbeta\ intensities and large $W({\rm WR})$ may e.g.\ be 
a flatter IMF. These are discussed in Sect.\ \ref{IMF}.

Regarding the signatures essentially attributed to WC stars 
(i.e.\ \niii + \Ciii, \civ) we note a shift in $W(\hbeta)$ 
between the observed and predicted equivalent widths (bottom panels 
of Fig.~\ref{fig_ew}, and Fig. \ref{fig_ew_004}). 
The shift looks as if WC stars appeared to early and/or did not
remain alive long enough. 
Possible explanations for this ``chronometer-shift'' are:
effects {\em (i), (b), (d)}, or variations in the burst duration, 
IMF, etc. (see Sect.~\ref{IMF}).
The uncertainties in the interpolation techniques determining the
number of WC stars ({\em (b)}) seem large enough to explain this
effect. 

For Tol 89 we measure an extraordinarily strong \civ\ emission 
($I(5808)/I(\hbeta) \sim 0.13$, $W(5808) \sim$ 12 \AA) which exceeds
the values shown Figures \ref{fig_ew} and \ref{fig_ew_004}. 
The other W-R emission lines do not
show any particularity compared to the other regions at $Z \sim 0.004$.
We have no simple explanation for this strong emission. 
Other observations are required to confirm the measurements for 
Tol 89.

\subsection{Age, duration and IMF of starbursts}
\label{IMF}

\begin{figure}[t]        
\centerline{\psfig{figure=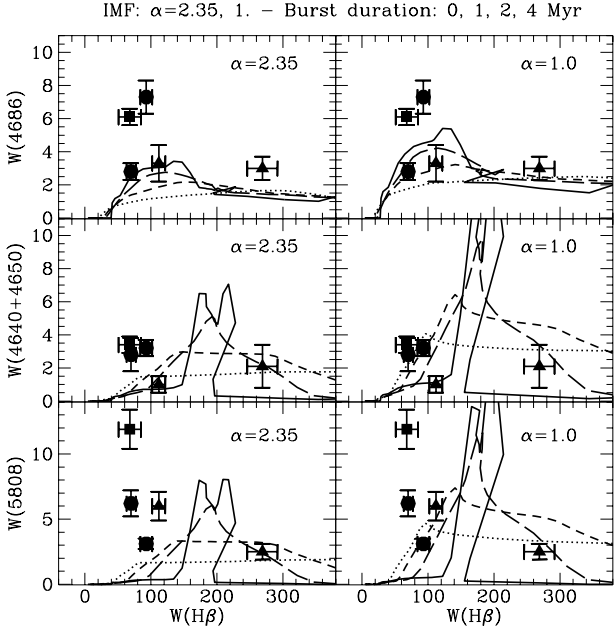,width=8.8cm}}
\caption{Observed and predicted W-R equivalent widths for 
objects with $Z \sim$ 0.004. Observations shown using the same
symbols as in Fig.\ \protect\ref{fig_ihb}. 
Models with different IMF slopes (left panels: $\alpha=2.35=$Salpeter;
right:  $\alpha=$ 1.0) and burst durations of $\Delta t=$ 0 
(``instantaneous'', solid lines), 1 (long dashed), 2 (short dashed), 
and 4 Myr (dotted) are shown. Discussion given in text
}
\label{fig_ew_004}
\end{figure}

The mere detection of W-R signatures in the integrated spectrum of galaxies 
reveals the presence of stars with ages generally between 1 and 8 Myr 
(Meynet, 1995; SV98), depending on the metallicity.
These predictions, based on single star evolutionary models, apply
for stars which have undergone ``normal'' mass loss.
The possibility of additional mass loss through Roche-lobe overflow
in a close binary system is not considered here 
\cite{CERVINOetal96,SV96,VANBEVERENetal97,SV98,C98}.
An independent age estimate is obtained from $W(\hbeta)$.
From our standard models at the appropriate metallicity, we obtain ages 
between 3 and 6 Myr, which are reported in Table~\ref{STARPOP} (column 2) 
for each starburst region.
As mentioned earlier these ages agree well with age range predicted for the 
W-R phase by the SV98 models at the appropriate metallicity and give
some support to $W(\beta)$--age relation.

Models with different burst durations are plotted in Fig.~\ref{fig_ew_004}.
These show that, within reasonable assumptions for the IMF slope,
the duration of star formation episode is limited to $\Delta t \la 2 - 4$ Myr.
From Fig.~\ref{fig_ew_004} and taking into account effect {\em (iii)} this
is a conservative limit; longer durations are clearly excluded from this
basis (see also Sect.\ \ref{s_r136}).
This quantitative estimate is
compatible with most studies of W-R galaxies which have so far clearly 
favoured instantaneous burst scenarios \cite{KS81,ARNAULTetal89,VC92,M95,S96}.

What can be said about the IMF slope?
As mentioned earlier the observations of the W-R features in NGC~3049, He~2-10,
and NGC~5253 are quite compatible with a Salpeter IMF. 
However, the large observed values of $W(4686)$ in NGC~3125-A and Tol 89, 
and $W(5808)$ in Tol 89 (representing in principle already lower limits 
due to {\em (iii)}) may require more W-R stars. This can be obtained
by invoking a flatter IMF, which increases the WR equivalent widths as shown 
in Fig.\ \ref{fig_ew_004} .
However, we note that intrinsicly the model predictions for \heiia\ are not 
very sensitive to changes of the IMF slope.
More IMF-sensitive are the features of WC stars, which descend on the average 
from more massive predecessors than WN stars and from a narrower range of 
initial masses.

Although the standard models show some deficiencies it is difficult to 
claim significant differences with respect to a Salpeter IMF for the 
following reasons:
1) The \heiia\ predictions are not very sensitive to the IMF slope. 
Furthermore the prediction of this line is uncertain (see {\em (c)}, and the
observed line  may also be contaminated by nebular emission 
(e.g.\ in NGC~3125-A, Sect.\ \ref{MASSPOP}).
2) Most of the WC signatures from the discrepant objects are within the 
range of the model predictions for a Salpeter IMF.
We therefore conclude that within the uncertainties our observations are 
compatible with a Salpeter IMF. 
Although we cannot exclude this possibility, no clear case requiring a 
significantly flatter IMF is found.
Much steeper IMF slopes may, however, not be compatible with our data.

Regarding the set of stellar tracks (cf.\ {\em (a)})
we note that the present results are only obtained for the stellar evolution
models of Meynet \etal\ \cite*{MEYNETetal94} adopting high mass loss rates.
Significantly lower equivalent widths of \heiia\ (approximately factor $\la 0.5$),
\niii + \Ciii\ ($\la 1/6$), and \civ\ ($\la 0.1$) are obtained with
the standard mass loss models due to the reduced WN and WC populations 
\cite{M95}.
As mentioned before the high mass loss models are favoured
by several independent comparisons with individual W-R stars and populations in 
the Local Group \cite{MM94}. This justifies their use when analysing
massive star populations in extra-galactic objects.

\begin{figure}[t]        
\centerline{\psfig{figure=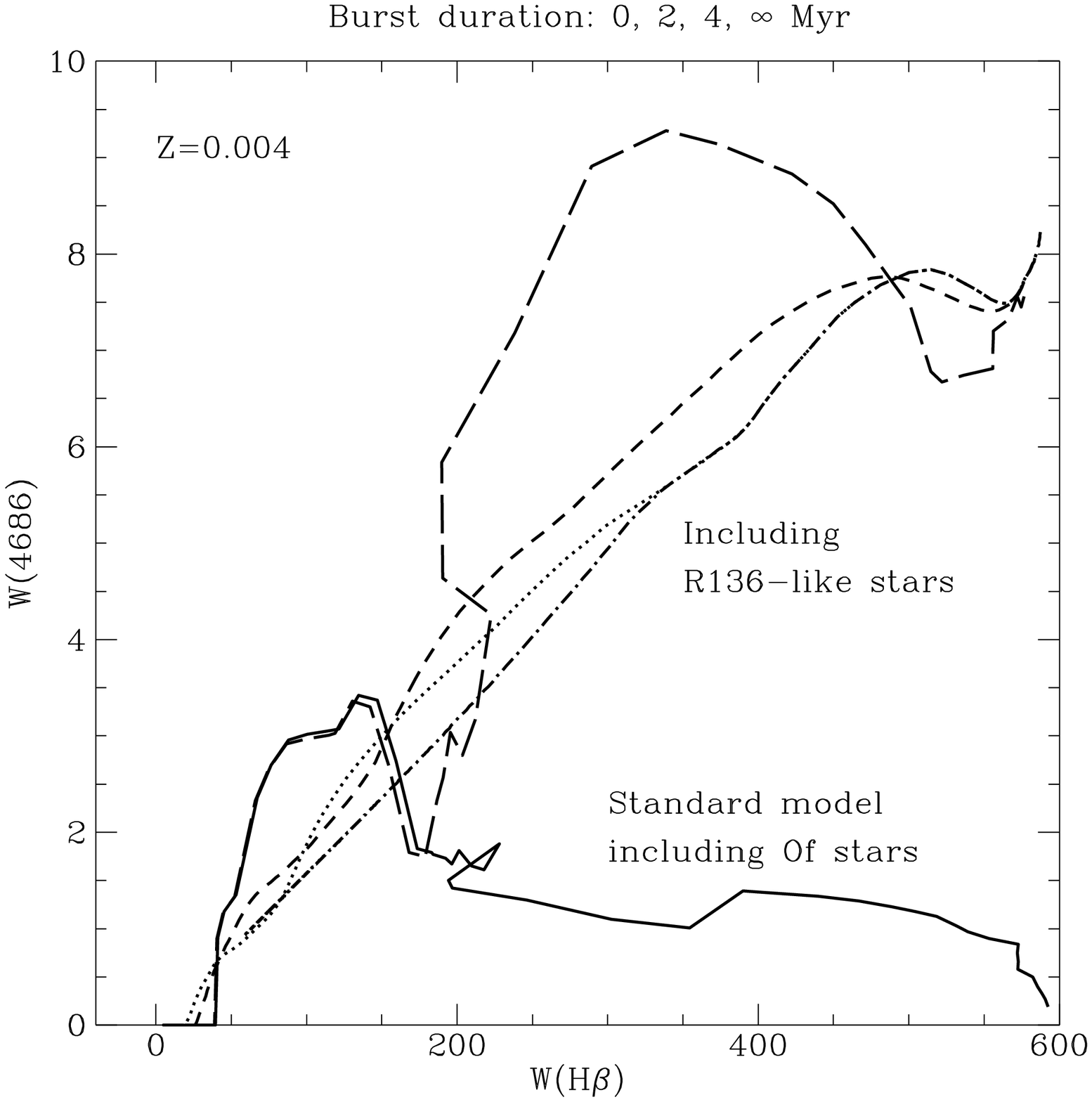,width=8.8cm}}
\caption{Predicted W-R equivalent widths (\protect\heiia) as a function of
$W(\hbeta)$ for the standard instantaneous burst model (solid, SV98),
and models including \protect\heiia\ emission from massive 
($M_{\rm ini} \ge$ 80 \protect\msun) main-sequence stars (``R136-like'' stars)
with varying burst durations: long-dashed (instantaneous burst), 
dashed ($\Delta t=$ 2 Myr), dotted ($\Delta t=$ 4 Myr), dashed-dotted
(continuous burst).
All models are for Z=0.004 and a Salpeter IMF.}
\label{fig_r136}
\end{figure}

\subsection{On the influence of R136 like stars on spectra of  W-R galaxies}
\label{s_r136}

We shall now briefly digress and investigate how robust the above results 
(especially the burst duration) are in view of recent studies of massive 
stars in R136 and the Galactic \hii\ region NGC~3603.
{\it HST} observations of these regions show the presence of very young 
($\la$ 2 Myr) stars with considerably strong \heiia\ emission (Drissen 
\etal, 1995; de Koter \etal, 1997).
Although their exact spectral types are still under debate (e.g.\ Crowther \& 
Dessart, 1998) quantitative analysis indicate that these stars are likely
massive hydrogen-burning stars which are less evolved than typical WN stars
(de Koter \etal, 1997; Crowther \& Dessart, 1998).
According to the criteria used to define W-R stars, traditional synthesis 
models would most likely miss to count such stars as W-R and hence their 
\heiia\ emission would not be included. If such stars are common in young 
star forming regions, the He~{\sc ii} emission predicted by the models 
would thus be underestimated. 
Could the presence of such emission line stars relieve the need to invoke
very short star formation timescales ?

To answer this question we investigated the effect of ``R136-like'' stars on 
the interpretation of W-R galaxy spectra, by performing test calculations 
(see Fig.~\ref{fig_r136}).
To account to first order for \heiia\ emission from fairly unevolved stars
(OIf and ``R136-like'' stars), we have assumed that all main-sequence stars 
with initial masses $M_{\rm ini} \ge$ 80 \msun\ (cf.\ de Koter \etal, 1997; 
Crowther \& Dessart, 1998) show a He~{\sc ii} emission with 
$L_{\rm 4686} \sim 1.7 \times 10^{36}$. 
This should provide an upper limit for the emission from ``R136-like'' 
stars\footnote{SV98 included a contribution 
of $L_{\rm 4686}=2.5 \times 10^{35}$ ergs s$^{-1}$ from main-sequence stars 
with low gravity. For the three most luminous stars of R136 and NGC~3603, 
the average 4686 luminosity is, however, larger: $L_{\rm 4686} \sim (1.8-2.2) 
\times 10^{36}$ ergs s$^{-1}$ (Crowther, 1997 and Drissen, 1997, private 
communications; de Koter 1996, private communication gives $L_{\rm 4686}$ lower 
by approximately a factor of 2) similar to WNL stars (see SV98).}.

Figure \ref{fig_r136} shows that, fairly independently of the star formation
history, the expected increase of \heiia\ due to the additional He\,{\sc ii} 
emitters is found only at $W$(\hbeta) $\ga 200$ \AA. 
At lower equivalent widths the differences with the ``standard model'' are
quite small (cf.\ Fig.~\ref{fig_ew_004}). If a general phenomenon, the existence 
of ``R136-like stars'' may thus help to explain the observations of \heiia\ only in 
regions with large \hbeta\ equivalent widths.
Our observations of W-R galaxies 
(see Fig.~\ref{fig_ew_004} and Table~\ref{DATA}) and those of VC92 are 
mostly found at $W$(\hbeta) $\la 100$ \AA. For these objects, the \heiia\ 
requires fairly short burst (see Fig.~\ref{fig_ew_004} and Fig.\ 2 of Schaerer 
1996). In any case ``R136-like'' stars cannot explain the observed \niii+\Ciii\ 
and \civ\ emission which also requires short burst durations as discussed above. 
Other independent arguments against extended star formation in
 the observed W-R galaxies are discussed below.

\section {Summary and conclusions}
The main result of the present paper is the unambiguous detection of
WC stars (indicated by broad \civ\ emission) in five previously known W-R galaxies, 
defined by broad \heiia\ emission mostly due to WN stars (Sect.\ \ref{W-RSTARS}).
We confirm the presence of WC stars in He 2-10 indicated by VC92.
With our four new detections (in NGC~3125, NGC~5253, Tol 89, and NGC~3049)
the total number of extragalactic objects known to harbour both WN and WC 
stars is now $\sim$ 19 (cf.\ Schaerer \& Contini 1998), which represent 
only $\sim$ 20\% of the total sample of W-R galaxies.
The relative weakness of \civ\ compared to \heiia\ ($I(5808)/I(4686) \la 0.5$
typically) and its larger width requires sufficiently high S/N ($\ga$ 40) to
be detected. As already pointed out by Schaerer \etal \cite*{SCHAERERetal97}
this explains most likely the non-detection in previous observations.

In all objects broad lines of \niii + \Ciii, \heiia, and \civ\ are measured
(Sect.\ \ref{W-RSTARS}).
A marginal detection of \heiib\ is found in two W-R regions; weak \ciii\ indicative
of late WC stars is found in the high metallicity W-R region of NGC~3049. 
From these emission lines we conclude that all W-R regions (except NGC~3049) contain 
a mixed population of WNL, and early WC and/or WO3-4 stars. This agrees well
with expectations (e.g.\ Maeder, 1991; SV98).

We have performed a detailed spatial analysis of the nebular and W-R emission
lines and the continuum light along the slit position in all our objects 
(Sect.\ \ref{SPATANAL}).
In He 2-10 and Tol~89 we found multiple peaks of nebular emission and a spatial 
offset between the main peak and the stellar continuum.
These structures are likely due to the existence of bubbles and loops in the
ISM powered by the kinetic energy released by massive stars (stellar winds
and/or SNe, see Sect.\ \ref{ENERGY}).
The spatial distribution of W-R stars follows closely the continuum and 
no significant distinction is found between WN and WC stars. 
The only exception is a bright \heiia\ peak with no continuum and nebular counterpart
found in NGC~5253 \cite{SCHAERERetal97}. Its origin is still unclear.

From the luminosity of the W-R signatures we have estimated the absolute number
of W-R stars of the different subtypes (Sect.\ \ref{MASSPOP}). 
For the regions whose \hbeta\ luminosities vary approximately up to a factor of 20
the total W-R star content varies from $\sim$ 30 to 1500.  
The estimated WC/WN number ratios (lower limits) are between 0.15 -- 0.65, with 
typical values between 0.2 -- 0.4 and no clear trend with metallicity. 
For our objects with metallicities $Z \sim 1/5$ \zsun, these values are larger than 
the observed WC/WN ratios in Local Group objects with similar $Z$ \cite{MJ98}. 
We argue that our WC/WN values are compatible with expectations for regions of 
short star formation. 
For He 2-10 and NGC~3049 the derived WC/WN ratio is below the trend given by 
Massey \& Johnson \cite*{MJ98}. This can have several explanations:
1) a short burst is observed at a particular time (quite low probability),
2) the number of WC stars is systematically underestimated (see Sect.\ \ref{MASSPOP}). 
The solution awaits new observations and quantitative analysis of ``WC+WN galaxies'' 
of different metallicities.

The detection of emission in the ``traditional'' blue W-R bump yields useful 
information about the age of the stellar population and the presence of massive 
stars (cf.\ Sect.\ \ref{MODELS}). 
Above this, detecting both WN and WC features provides a considerable 
improvement for the following reason:
WC stars are strongly evolved descendents of massive stars revealing He-burning
products on their surface; prior to this phase these objects are WN stars 
showing H-burning products (e.g.\ Maeder \& Conti 1994).
Given this WN $\rightarrow$ WC sequence, which is expected to be followed
only by the most massive WN stars, it is clear that the predictions of WC/WN 
populations are particularly sensitive to the evolutionary scenario and burst 
parameters (e.g.\ IMF, burst duration).
In Sect.\ \ref{MODELS} we have exploited this fact by performing detailed 
comparisons of three different observational W-R signatures from WC and WN 
stars with predictions from the recent synthesis models of SV98.

The comparisons of the observed equivalent widths and line intensities 
relative to \hbeta\ with models do not all show a simple picture. 
The most important effects and uncertainties which may affect such a 
comparison have been amply discussed in Sect.\ \ref{MODELS}.
Some W-R signatures in few regions exceed the predicted relative intensities 
and/or equivalent widths.
The majority of the observed quantities (\niii+\Ciii, \heiia, \civ, and \hbeta) 
can, however, be reproduced reasonably well by the SV98 models with a Salpeter IMF. 
Although some W-R lines may indicate a flatter IMF in some regions,
no clear case requiring a significantly flatter IMF is found.
Much steeper IMF slopes may, however, not be compatible with our data.
These results are in agreement with other studies of similar objects 
(e.g.\ Mas-Hesse \& Kunth 1991, 1998; Schaerer 1996; Leitherer 1998).

In order to reproduce the W-R lines, young populations with short durations 
of star formation are required. From our quantitative modeling, we find a
conservative limit for burst durations of typically $\Delta t \la 2 - 4$ Myr. 
A simple experiment shows that this result holds even if very young massive 
emission line stars such as found in R136 and NGC 3603 are common in our 
observed star-forming regions.
Such short star formation timescales can be understood if the light
observed in our W-R regions comes from one or few individual 
compact regions, such as the super-star clusters frequently identified 
on high resolution {\it HST} images (e.g.\ Conti \& Vacca, 1994; 
Meurer \etal, 1995).
The finding of young and short bursts is supported by other independent 
constraints from \hii\ galaxies which have very similar properties
to our W-R galaxies, such as the observed spread of $W$(\hbeta) and 
variations of nebular line ratios \cite{SL96}.

Last, but not least, we also note from our comparison with starburst models
that it is not possible to reproduce the observed WN and WC signatures adopting 
evolutionary models using standard mass loss rates. The high mass loss
models of Meynet \etal \cite*{MEYNETetal94} are clearly favoured from
our comparison (see also Maeder \& Meynet, 1994).
A similar conclusion was obtained from comparisons of new evolutionary
tracks with the observed W-R population in the extremely metal-poor galaxy 
I Zw 18 \cite{Detal98}, showing the usefulness of such observations
to constrain evolutionary scenarios in environments (i.e.\ extreme
metallicities) inaccessible in the Local Group.

Our successful finding of WC stars in W-R galaxies 
opens the door to new systematic studies of massive star populations 
in starbursts. In addition to the subjects addressed in the present 
paper the study of WC stars in starburst regions is of interest for 
several other reasons:
{\em 1)} WC stars may contribute to a harder ionizing spectrum which has 
implications on the observed nebular properties (e.g.\ Terlevich \& Melnick 
1985, Schaerer 1996).
{\em 2)} Regions which harbour WC stars during their lifetimes are expected
to produce ejecta of significantly different composition (cf.\ Maeder 1992).
Our understanding of stellar evolution, chemical evolution and the starburst
phenomenon should greatly benefit from such future studies.

\begin{acknowledgements}
We thank Bill Vacca for providing us with spectra of W-R galaxies for
comparisons and the referee, Peter Conti, for his positive comments.
We also benefited from fruitful discussions with Roger Coziol, Miguel 
Mas-Hesse, Georges Meynet, and Maximilien Pindao and from comments on 
earlier versions of the manuscript.
DS acknowledges support from the Swiss National Foundation of Scientific
Research.
This research has made use of the Lyon-Meudon Extragalactic Database 
(LEDA) supplied by the LEDA team at the CRAL-Observatoire de Lyon (France).

\end{acknowledgements}

\end{document}